\documentclass[a4paper,11pt]{article}
\usepackage{authblk}
\usepackage[parfill]{parskip}
\usepackage{color}
\usepackage{bm}
\usepackage{amsmath, amsthm, amssymb}
\usepackage{mathtools}
\usepackage{pdfpages}
\usepackage{textpos}
\usepackage{graphicx}
\usepackage{epstopdf}
\usepackage{booktabs}
\usepackage{longtable}
\usepackage{pdflscape}
\usepackage{soul}
\usepackage[dvipsnames]{xcolor}
\usepackage{setspace} 

\newtheorem{prop}{Proposition}

\usepackage[backend=biber,style=authoryear-comp,natbib=true,maxcitenames=3,giveninits=true, uniquename=false,sortcites=false,doi=false,isbn=false,maxbibnames=99, uniquelist=false]{biblatex}
\def\biz{\begin{itemize}}
\def\eiz{\end{itemize}}
\def\ben{\begin{enumerate}}
\def\een{\end{enumerate}}

\def\Xb{\bm{X}}
\def\yb{\bm{y}}
\def\eb{\bm{e}}

\def\bb{\bm{b}}
\def\yhat{\hat{\yb}}
\def\ytilde{\tilde{\yb}}
\def\bhat{\hat{\bb}}
\def\btilde{\tilde{\bb}}
\def\Sb{\bm{S}}
\def\Yb{\bm{Y}}
\def\Cb{\bm{C}}

\def\Wb{\bm{W}}
\def\Jb{\bm{J}}
\def\Eb{\bm{E}}
\def\eb{\bm{e}}
\def\Zb{\bm{Z}}
\def\Bb{\bm{B}}
\def\Db{\bm{D}}
\def\Gb{\bm{G}}

\title{Hierarchical forecasting: The role of information}

\author{Minh Nguyen}
\author{Farshid Vahid}
\author{Shanika L Wickramasuriya\footnote{Corresponding author. Shanika L Wickramasuriya, Department of Econometrics and Business Statistics, Monash University, Caulfield East, VIC 3145, Australia. Email address: \texttt{shanika.wickramasuriya@monash.edu}.}}
\affil{Department of Econometrics and Business Statistics, Monash University, Australia}
\date{\today}

\usepackage{hyperref}
\definecolor{darkblue}{rgb}{0,0,.6}
\hypersetup{
  citecolor = 0 0 0,
  colorlinks = true,
  linkcolor = darkblue,
  urlcolor = blue,
  citecolor = darkblue,
  breaklinks = true
}

\providecommand{\keywords}[1]{\textbf{Keywords: } #1}

\begin{document}

\maketitle

\begin{abstract}
In hierarchical forecasting, the process of forecast reconciliation transforms a set of ``base'' or ``raw'' forecasts, which do not satisfy the hierarchical aggregation constraints in the real data, into a set of ``coherent'' forecasts, which do satisfy those constraints. The academic literature provides ample simulation evidence and real-world examples demonstrating the value of forecast reconciliation in improving forecasts of hierarchical time series. This improvement is attributed to the imposition of aggregation constraints. However, this evidence is derived from base forecasts, each generated using a distinct information set, usually the univariate information set corresponding to each time series. Since reconciliation algorithms combine forecasts, it is difficult to determine the extent to which the improvement is due to the imposition of constraints versus the combination of information carried by different forecasts.

In this paper, we demonstrate that when base forecasts are based on different information sets and historical data are available, there is scope for improving these forecasts by combining the information that each one carries, even when they are already coherent. We propose a new method, called the \textbf{information combination} (IComb) method, which combines the information content of forecasts during the reconciliation process. The method is regression-based and can be implemented using existing penalised regression packages. We provide simulation evidence to illustrate the role of information sets, as distinct from the role of aggregation constraints, in forecasting hierarchical time series. Finally, we apply our method to datasets previously used in the literature and demonstrate that it achieves superior results compared to traditional approaches.
\end{abstract}


\section{Introduction}{\label{sec:intro}}

The seminal work of Sir Richard Stone on developing a system of national accounts, in particular his work with David Champernowne and James Meade \citep{Stone_1942} on providing a measure of precision for components of national accounts, was an important advance in the theory and practice of estimation under aggregation constraints. \citet{Byron1978} provides a summary of Stone's work and reformulates the problem as a constrained optimisation problem. \textit{Hierarchical forecasting} is a relatively recent development in this area that has become increasingly important given the needs of global value chains and multinational retailers. There is, however, a very important difference between providing best coherent estimates for the components of national accounts whose true values will never be known, and providing coherent forecasts for a set of variables whose values will become known in the future usually without any measurement error. While the hierarchical forecasting literature has benefited from the body of research that built on the work of Richard Stone and his co-authors, it has exploited the advantage that the accuracy of forecasts can be evaluated \textit{ex post}.  

In hierarchical forecasting, the process of forecast reconciliation transforms a set of ``base'' forecasts (also referred to as ``raw'' forecasts in the literature), which do not satisfy the hierarchical aggregation constraints in the real data, into a set of ``coherent'' forecasts, which do satisfy those constraints. The reconciliation algorithms are mainly variants of either the ordinary least squares algorithm of \citet{HyndmanEtAl2011}, which can be applied when there is no historical data to estimate the accuracy of base forecasts, or the minimum trace algorithm of \citet{WicEtAl2019}, which can be applied when there is sufficient historical data to estimate the accuracy of base forecasts.

The value of forecast reconciliation in improving hierarchical time series forecasts is attributed to the value of imposing the aggregation constraints. However, this evidence is derived from base forecasts each generated using a distinct information set, usually the univariate information set corresponding to each time series. Since reconciliation algorithms combine forecasts, it is difficult to identify to what extent the improvement is due to imposing a true constraint or to combining the information carried by different forecasts.

In this paper, we consider the importance of information sets in hierarchical forecasting. We examine their role in the existing results and provide fresh insights on where and when the existing algorithms succeed or fail. We also propose a new method, which we call the ``information combination'' (IComb) method, and provide some new results and recommendations for forecast reconciliation. It is worth noting that all the aforementioned methods are applicable not only to aggregation constraints but also to any known set of linear constraints.

We consider a hierarchical $m \times 1$ vector time series $\yb_t $, in which the top $m-n$ components are aggregates of its bottom $n$ components, which are denoted by $\bb_t$. The $m \times n$ aggregation matrix $\Sb$ transforms $\bb_t$ to $\yb_t $, i.e.,
\begin{equation}
	\yb_t = \Sb \bb_t \text{, for all }t.
\end{equation}  
The aggregation matrix has the following structure:
\begin{equation}
	\Sb = \begin{bmatrix}
		\Cb \\ \bm{I}_n
	\end{bmatrix},
\end{equation} 
where $\Cb$ is the matrix of aggregation constraints, and $\bm{I}_n$ is the identity matrix of size $n$.
We also define the $m \times (m-n)$ matrix $\Sb_\perp$ as
\[ 
\Sb_\perp = \begin{bmatrix}
	\bm{I}_{m-n} \\ -\Cb'
\end{bmatrix},
\]
where $(\cdot)'$ denotes the matrix transpose. This matrix\footnote{This is the matrix $\bm{U}$ in \citet{WicEtAl2019}. We prefer $\Sb_\perp$ because it makes clear that this is a basis for the orthogonal complement of the column space of $\Sb$.} has the property that $\Sb_\perp' \Sb = \bm{0}$. The hierarchical structure of $\yb_t$ imposes $m-n$ linear restrictions at any time $t$ given by
\begin{equation}
	\Sb_\perp' \yb_t = \bm{0}.
\end{equation}  
The vectors $\yhat_{T+h|T}$ and $\bhat_{T+h|T}$ denote the ``base'' forecasts (forecasts provided, possibly by others, which may not satisfy the aggregation constraints embodied in $\Sb$) of $\yb_{T+h}$ and $\bb_{T+h}$ at time $T$ respectively, and $\ytilde_{T+h|T}$ and $\btilde_{T+h|T}$ denote the reconciled forecasts of $\yb_{T+h}$ and $\bb_{T+h}$ at time $T$, respectively, so that,
\begin{equation*}
	\ytilde_{T+h|T} = \Sb \btilde_{T+h|T} \text{, for }h=1,2,\ldots.
\end{equation*}

\section{The role of information and a new information combination approach to reconciliation}

Let $\mathcal{I}_T$ denote the information set, that is, all observations available at time $T$, i.e., $\mathcal{I}_T = \left\lbrace \yb_1,\ldots,\yb_T \right\rbrace$. It is well known that the best forecast of a random variable given an information set, in the mean squared error sense, is its conditional expectation. In the context of forecasting a hierarchical vector time series, this implies that the best forecast of $\yb_{T+h}$ given $\mathcal{I}_T$, in the aggregate MSE sense (sum of the mean squared errors of each element), is $E\left( \yb_{T+h} \mid \mathcal{I}_T \right)$. Given the linearity of the expectation operator, we note the following proposition:

\begin{prop}\label{p:ce}
	The conditional expectation function of a hierarchical time series given any information set satisfies the aggregation constraints that the actual time series satisfies. 
\end{prop}

The proof is straightforward. The linearity of the expectation operator implies
\[ E(\yb_{t} \mid \Omega) = \Sb E(\bb_t \mid \Omega ) \text{ for any }t \text{ and }\Omega,\]
which in turn implies $\Sb_\perp' E(\yb_{t} \mid \Omega) = \bm{0}$.

Since this holds for any information set, it also holds unconditionally. More importantly, since it holds for any $t$ and $\Omega$, it implies that
\[
\Sb_\perp' E(\yb_{T+h} \mid \mathcal{I}_T) = \bm{0}.
\]   
This means that if the conditional expectation functions of all elements of $\yb_{T+h}$ were known and used to produce forecasts, those forecasts would automatically satisfy the aggregation constraints. Of course, the conditional expectation functions are unknown, and research on hierarchical forecasting is motivated by the realistic scenario that the base forecast of each component is provided by possibly different forecasters (or algorithms) and these forecasts do not satisfy the aggregation constraints. This motivates the research question of how to combine these forecasts to minimise the aggregate MSE across all components.

When there is no a priori knowledge of the accuracy of the forecasts for any component and no historical forecasts are available to assess this accuracy empirically, \citet{HyndmanEtAl2011} propose an elegant and simple solution that produces a set of reconciled forecasts that satisfy the aggregation constraints and are closest in Euclidean distance to the vector of base forecasts. Specifically, they propose
\begin{equation}
	\ytilde_{T+h|T}^{\text{OLS}} = \Sb(\Sb'\Sb)^{-1}\Sb'\yhat_{T+h|T}.
\end{equation}
This solution is easy to implement and always feasible. \citet{panagiotelis2021forecast} provide a geometric interpretation and show that the OLS reconciled forecasts are at least as good as the base forecasts in the mean squared error sense. As shown in many papers \citep[see, among others,][and we will demonstrate in an example below]{DiFG2024,HollymanEtAl2021}, this solution provides a weighted average of the direct and indirect forecasts of each element using a set of fixed weights that ensure the combined forecasts satisfy the hierarchical aggregation constraints. The seminal work of \citet{BaGr1969} provides a further reason to believe that reconciled forecasts can have lower MSE than the base forecasts, due to the fact that they are formed by combining forecasts. 

When a sufficiently long history of forecasts and actual observations is available, \citet{WicEtAl2019} propose a reconciliation procedure assuming that base forecasts are conditionally unbiased, i.e.,
\begin{equation}
E(\yb_{T+h}-\yhat_{T+h|T} \mid \mathcal{I}_T) = \bm{0}, \label{eq:unbiased}	
\end{equation}
which implies
\[
E(\yhat_{T+h|T} \mid \mathcal{I}_T)=E(\yb_{T+h} \mid \mathcal{I}_T).
\]
This is a very strong requirement because it implies that 
\[
\yhat_{T+h|T} = E(\yb_{T+h} \mid \mathcal{I}_T) + \bm{\varepsilon},
\]
where $E(\bm{\varepsilon} \mid \mathcal{I}_T) = \bm{0}$. This means that all useful information in the history of \textit{all} components of the hierarchy has already been embodied in the base forecasts, and their deviation from the conditional expectation function cannot be learned from $\mathcal{I}_T$. Under this assumption, all gains from forecast reconciliation will be due to the aggregation constraints. The role of history reduces to estimating the $m \times m$ positive definite base forecast error variance-covariance matrix 
\[\hat{\bm{W}}_h = \frac{1}{T-h+1} \sum_{t=0}^{T-h} \hat{\bm{e}}_{t+h|t} \hat{\bm{e}}_{t+h|t}',\]
where $\hat{\bm{e}}_{t+h|t} = \yb_{t+h} - \yhat_{t+h|t}$, and then forming a weighted least squares regression of $\yhat_{T+h|T}$ on the columns of $\Sb$ with $\hat{\bm{W}}_h^{-1/2}$ as weights, leading to
\begin{equation}
\ytilde_{T+h|T}^{\text{MinT}} = \Sb \left(\Sb'\hat{\bm{W}}_h^{-1}\Sb \right)^{-1}\Sb'\hat{\bm{W}}_h^{-1} \yhat_{T+h|T}. \label{eq:mint}
\end{equation}
This is the MinT (minimum trace) sample reconciliation method proposed by \citet{WicEtAl2019}. In the rest of the paper, we refer to this approach as ``MinT sample". They formulate the forecast reconciliation problem in terms of finding the $n \times m$ matrix $\Gb$ in $\ytilde_{t+h|t} = \Sb \Gb \yhat_{t+h|t}$ such that the trace of the variance-covariance matrix of the reconciled forecast errors is minimised, subject to $\Sb \Gb \Sb = \Sb$, which is a sufficient condition\footnote{The reason why this condition is sufficient but not necessary is explained in Appendix \ref{app:A}.} for conditional unbiasedness of the forecasts.

The performance of these reconciliation methods, and some of their variants, has been studied in the literature via Monte Carlo simulations and hypothetical scenarios with real data. However, in most of those exercises, the base forecasts are not conditionally unbiased in the sense of equation (\ref{eq:unbiased}). In almost all of such explorations, the base forecast for each component is produced from a univariate time series model based only on the history of that component. That is, the forecast of $y_{i,T+h}$ is not based on all information in $\mathcal{I}_T$, but only on a subset, namely $\mathcal{I}_{i,T} = \left\lbrace y_{i,1}, y_{i,2}, \ldots, y_{i,T} \right\rbrace $. Even if we had the ideal forecast of $y_{i,T+h}$ in the MSE sense based on $\mathcal{I}_{i,T}$, i.e., $\hat{y}_{i,T+h|T} = E(y_{i,T+h} \mid \mathcal{I}_{i,T})$, the forecast error would, in general, not have zero expectation conditional on $\mathcal{I}_T$. The only exception is when none of the components are Granger-caused by any other component, i.e., when all are actually generated by univariate time series models. This condition corresponds to a single point in the simulation design in \citet{WicEtAl2019}, and is very unlikely to hold in any real data example. 

The question, then, is how to approach reconciliation in a more realistic situation where base forecasts may be based on different information sets. We abandon the assumption that base forecasts are unbiased in the sense of equation (\ref{eq:unbiased}), and instead ask how the information in the base forecasts might be combined to produce better forecasts that also satisfy the aggregation constraints. To answer this, we define $\bm{Y}$ as the $T \times m$ matrix 
\[
\bm{Y} = \begin{bmatrix}
	\yb_1' \\
	\yb_2' \\
	\vdots \\
	\yb_T'
\end{bmatrix}
\]
and let $\bm{X}$ be any $T \times k$ matrix, and we propose the following:

\begin{prop}
	Assume $\bm{X}$ has full column rank. In the multivariate regression of $ \bm{Y}$ on $\bm{X}$, 
	\[
	\bm{Y} = \bm{X} \bm{B} + \bm{E},
	\]  
	the ordinary least squares estimator of the coefficient matrix $\hat{\Bb} = (\Xb'\Xb)^{-1}\Xb'\Yb$ satisfies
	$\Sb_\perp ' \hat{\Bb}' = \bm{0}$. \label{prop:ols}
\end{prop}

The proof follows from the fact that $\Sb_\perp ' \bm{\hat{B}}' = \Sb_\perp '  \Yb' \Xb (\Xb'\Xb)^{-1}$ and $\Sb_\perp '  \bm{Y}' = \bm{0}$, given the definition of $\bm{Y}$. Also note that each column of $\bm{\hat{B}}$ is the least squares estimate of the regression of the corresponding column of $\bm{Y}$ on $\bm{X}$.

This leads us to start with the following reconciled forecast, aimed at combining the information content of base forecasts:
\begin{equation}
	\ytilde_{T+h|T}^{\text{IComb}} = \left(\bm{\hat{B}}_h^{\text{IComb}}\right)' \yhat_{T+h|T}, \label{eq:base}
\end{equation}
where the $i$-th column of $\bm{\hat{B}}^{\text{IComb}}_h$ is the OLS estimate of the parameters of a regression of $y_{i,t+h}$ on $\yhat_{t+h|t}$, using all observations and base forecasts up to time $T$. If unbiasedness in the conventional sense (zero unconditional expectation of the forecast errors) is desirable, an intercept can be added to these regressions. This is similar to the regression-based forecast combination procedure suggested by \citet{GraRam1984}, except that here the goal is to combine the information content of forecasts of different variables, rather than multiple forecasts of the same variable. We call this the ``information combination'' (IComb) reconciliation method. 

There are obvious issues with this proposed forecast reconciliation method. Even when the base forecasts are not coherent, they are likely to be nearly collinear, which makes the estimated parameters in these least squares regressions very imprecise. Moreover, when $m$ (or $k$) is large relative to $T$, these estimates again become highly imprecise. Hence, in practical applications, this reconciliation method may not improve forecast accuracy. Fortunately, there is a practical solution motivated by the following proposition:
\begin{prop}
	In the multivariate regression of $ \Yb $ on $\Xb$, 
	\[
	\bm{Y} = \bm{X} \bm{B} + \bm{E},
	\]  
	the ridge regression estimator of the coefficient matrix $\bm{\hat{B}}^{\nu} = \left(\bm{X}'\bm{X} + \nu \bm{\Lambda}_k\right)^{-1}\bm{X}'\bm{Y}$, where $\nu > 0$ is the shrinkage parameter and $\bm{\Lambda}_k$ is any $k \times k$ diagonal matrix with positive diagonal elements, satisfies
	$\Sb_\perp ' \left(\bm{\hat{B}}^{\nu}\right)' = \bm{0}$.  \label{prop:ridge}
\end{prop}
 
The proof is similar to the proof of proposition \ref{prop:ols}. When the norms of columns of $\Xb$ are of the same order of magnitude, then $\bm{\Lambda}_k=\bm{I}_k$ is a good choice, and it shrinks OLS coefficients to zero. When these columns have very different scales, as is almost always the case in hierarchical time series, then $\bm{\Lambda}_k = \Db$, where $\Db$ is a diagonal matrix with the same main diagonal as $\Xb'\Xb$, makes better sense and shrinks the standardised OLS coefficients to zero. 

This leads us to recommend the following reconciliation method, which we call the ``information combination with shrinkage'' (ICombS) method. This method combines the information content of base forecasts, makes them coherent and is scalable:
\begin{equation}
	\ytilde_{T+h|T}^{\text{ICombS}_{\nu}} = \left(\bm{\hat{B}}_h^{\text{ICombS}_{\nu}}\right)' \yhat_{T+h|T}, \label{eq:ridge}
\end{equation}
where the $i$-th column of $\bm{\hat{B}}_h^{\text{ICombS}_{\nu}}$ is the estimated vector of parameters of a ridge regression of $y_{i,t+h}$ on scaled $\yhat_{t+h|t}$ (i.e., using $\bm{\Lambda}_k = \Db$) with shrinkage parameter $\nu$, using all observations and base forecast up to time $T$. It also makes sense to center all observations and then run the ridge regression to make the predictions to shrink towards historical means rather than towards zero. Since the sample means also satisfy the aggregation constraints, this does not affect the coherency of the resulting forecasts. In cases where components of $\yb_{t}$ are non-stationary, one can also shrink the combinations forecasts towards random walks with drifts, and since $\yb_t$ satisfies the aggregation constraints for all $t$, $\yb_t-\yb_{t-1}$ will also satisfy these constraints. Note that the shrinkage parameter must be the same for all $i$, i.e., across all $m$ ridge regressions. The shrinkage parameter can be set by the rule explained in \citet{WicEtAl2019}, or via cross-validation using the \texttt{glmnet} package \citep{glmnet}. In this paper, we determine the optimal $\nu$ by rolling-window cross-validation, minimising the mean squared reconciled forecast error in the validation set; the details are provided in Section \ref{sec:empirical}.

From Propositions 2 and 3, it is clear that scaling the variables in $\bm{X}$ does not alter the results. However, the target variables in $\bm{Y}$ also have very different orders of magnitude, since aggregates have larger magnitudes than their components. It may appear that scaling the target variables will destroy the results in Propositions 2 and 3 because the scaled variables will no longer satisfy the aggregation constraints. While it is true that they will not satisfy the original aggregation constraints, they will satisfy another set of linear constraints, which the information combination forecasts (with or without shrinkage) will also satisfy. Hence, when the information combination forecasts are transformed back to the original scale, they will also satisfy the original aggregation constraints. This is stated formally in the following proposition: 
\begin{prop}
    Let $\bm{Y}^* = \bm{Y}\hat{\bm{\Sigma}}_y^{-1/2}$, $\bm{X}^* = \bm{X}\hat{\bm{\Sigma}}_x^{-1/2}$ and $\bm{B}^* = \hat{\bm{\Sigma}}_x^{1/2}\bm{B}\hat{\bm{\Sigma}}^{-1/2}_y$, where $\hat{\bm{\Sigma}}_x$ and $\hat{\bm{\Sigma}}_y$ are diagonal matrices whose elements are given by the variance of each column in $\bm{X}$ and $\bm{Y}$, respectively.
	In the multivariate regression of $ \Yb^* $ on $\Xb^*$, 
	\[
	\bm{Y}^* = \bm{X}^* \bm{B}^* + \bm{E}^*,
	\]  
	the ridge regression estimator of the coefficient matrix $\bm{\hat{B}}^{*\nu} = \left(\bm{X}^{*'}\bm{X}^* + \nu \bm{\Lambda}_k\right)^{-1}\bm{X}^{*'}\bm{Y}^*$ and $\hat{\bm{B}}^\nu = \hat{\bm{\Sigma}}_x^{-1/2}\bm{\hat{B}}^{*\nu}\hat{\bm{\Sigma}}_y^{1/2}$, where $\bm{\Lambda}_k$ is any $k \times k$ diagonal matrix with positive diagonal elements, satisfies
	$\Sb_\perp ' \left(\bm{\hat{B}}^{\nu}\right)' = \bm{0}$.  \label{prop:ridge-scaled}
\end{prop}
The proof is based on the fact that $\Yb^*$ satisfies an alternative set of linear restrictions given by $\Sb_\perp '\hat{\bm{\Sigma}}_y^{1/2} (\Yb^*) '=\bm{0}$, which, by Proposition 3, implies $\Sb_\perp '\hat{\bm{\Sigma}}_y^{1/2} \left(\bm{\hat{B}}^{*\nu}\right) '=\bm{0}$. Therefore, $\Sb_\perp ' \hat{\bm{\Sigma}}_y^{1/2} \left(\bm{\hat{B}}^{*\nu}\right)'\hat{\bm{\Sigma}}_x^{-1/2} = \Sb_\perp ' \left(\bm{\hat{B}}^{\nu}\right)' = \bm{0}$.

It must be clear from Propositions \ref{prop:ols}--\ref{prop:ridge-scaled} that, in these multivariate regressions, each equation should include the same set of regressors. However, they need not be the forecasts of all $m$ components of the hierarchical time series. If the base forecasts of some components are so poor that, after having the other forecasts, they add no additional information to the pool, including them only adds noise and is likely to degrade forecast accuracy. Hence, one might wish to use a selection method such as the group lasso \citep{YuaLim2006} or the group subset \citep{ThoVah2024} instead of the ridge method. In fact, as we show in the following proposition, the multivariate lasso can be used for this purpose, which means that the \texttt{mgaussian} option of the \texttt{glmnet} package can be used to construct the information combination with shrinkage estimator.
\begin{prop}
	Let $\bm{X}, \bm{Y}$, and $\bm{B}$ be defined as in Proposition \ref{prop:ridge}. In the multivariate regression of $ \Yb $ on $\Xb$, 
	\[
	\bm{Y} = \bm{X} \bm{B} + \bm{E},
	\]  
	the multivariate lasso estimator of the coefficient matrix satisfies
	$\Sb_\perp ' \left(\bm{\hat{B}}^{\text{ml}}\right)' = \bm{0}$. \label{prop:ml}
\end{prop}

Interestingly, similar to Proposition \ref{prop:ridge}, the result generalises to the case where the dependent and independent variables are scaled, which we state formally in the following proposition:
\begin{prop}
	Let $\bm{X}^*, \bm{Y}^*$, and $\bm{B}^*$ be defined as in Proposition \ref{prop:ridge-scaled}. In the multivariate regression of $ \Yb^* $ on $\Xb^*$, 
	\[
	\bm{Y}^* = \bm{X}^* \bm{B}^* + \bm{E}^*,
	\]  
	the multivariate lasso estimator of the coefficient matrix, $\hat{\bm{B}}^{*\text{ml}}$, satisfies
	$\Sb_\perp ' \left(\bm{\hat{B}}^{\text{ml}}\right)' = \bm{0}$, where $\hat{\bm{B}}^{\text{ml}} = \hat{\bm{\Sigma}}_x^{-1/2}\hat{\bm{B}}^{*\text{ml}}\hat{\bm{\Sigma}}_y^{1/2}$. \label{prop:ml-scaled}
\end{prop}

The proof of Proposition \ref{prop:ml-scaled} is provided in Appendix \ref{app:mlasso}, while the proof of Proposition \ref{prop:ml} follows as a special case of Proposition \ref{prop:ml-scaled}. As mentioned above, the information combination forecast produced with multivariate lasso may exclude poor-quality forecasts of some components altogether, replacing them with coherent forecasts derived from the forecasts of the remaining components. We call this the ``information combination with lasso'' (ICombL) method.

\subsection{Positioning IComb within the existing forecast reconciliation approaches}
\label{sec:relationships}

Let us consider a simple hierarchy with two bottom-level series summing to a single top-level series, i.e., $m=3$ and $n=2$. That is, the vector time series $\yb_t$ has three elements, with $y_{1,t} = y_{2,t}+y_{3,t}$ for all $t$. The $\Sb$ matrix in this case is:
\begin{equation*}
	\Sb = \begin{bmatrix}
		1 & 1 \\
		1 & 0 \\
		0 & 1
	\end{bmatrix} \Rightarrow \Sb_\perp = \begin{bmatrix*}[r]
	1  \\
	-1  \\
	-1 
	\end{bmatrix*}
\end{equation*}
leading to the OLS reconciliation transformation:
\begin{align*}
	\ytilde_{T+h|T}^{\text{OLS}} &= \Sb(\Sb'\Sb)^{-1}\Sb'\yhat_{T+h|T} \\
	 &= \begin{bmatrix*}[r]
		2/3 & 1/3 & 1/3\\
		1/3 & 2/3 & -1/3\\
		1/3 & -1/3 & 2/3
	\end{bmatrix*} \yhat_{T+h|T},
\end{align*}
which means (suppressing the conditioning set to simplify notation)
\begin{align*}
	\tilde{y}_{1,T+h}^{\text{OLS}} &= \frac{2}{3} \hat{y}_{1,T+h} + \frac{1}{3} (\hat{y}_{2,T+h}+\hat{y}_{3,T+h}) \\
	\tilde{y}_{2,T+h}^{\text{OLS}} &= \frac{2}{3} \hat{y}_{2,T+h} + \frac{1}{3} (\hat{y}_{1,T+h}-\hat{y}_{3,T+h}) \\
	 \tilde{y}_{3,T+h}^{\text{OLS}} &= \frac{2}{3} \hat{y}_{3,T+h} + \frac{1}{3} (\hat{y}_{1,T+h}-\hat{y}_{2,T+h}). 
\end{align*}
The OLS reconciliation is therefore a weighted average of the direct and indirect forecasts of each component, with fixed $(2/3,1/3)$ weights. The MinT and IComb reconciliation methods determine these weights based on the observed data prior to time $T$. The MinT method requires that the final transformation be a projection onto the column space of $\Sb$, whereas the IComb method relaxes all restrictions except that the transformed vector must lie in the range of $\Sb$. As such, IComb can accommodate adding a constant or other information at the reconciliation stage.

It may appear that IComb is simply an alternative way of motivating the MinT sample reconciliation method. We find it instructive to explore the difference between these two methods more rigorously.

In Appendix \ref{app:MinT}, we show that the MinT sample can be written as (the reference to the conditioning set is suppressed to simplify notation)
\begin{equation*}
	\ytilde_{T+h} = \Sb \Gb \yhat_{T+h} = \Sb \ytilde_{T+h}^{bot}
\end{equation*}
and
\begin{equation*}
	\ytilde_{T+h}^{bot} = \yhat_{T+h}^{bot} + \bm{\Psi}' (\yhat^{top}_{T+h} - \Cb \yhat^{bot}_{T+h}),
\end{equation*}
where the superscript $bot$ denotes the $n$ bottom-level components of a vector, and the $i$-th column of $\bm{\Psi}$ contains the estimated parameters of the regression of the base forecast error in the $i$-th component of $\yb_t^{bot}$ on the $m-n$ forecast incoherencies $\yhat_t^{top} - \Cb \yhat_t^{bot}$, using all available data up to time $T$. This makes perfect sense if the assumption of conditional unbiasedness is correct. Under that assumption, there is no useful information in the forecasts of other variables, and the only relevant information lies in the incoherencies.

In contrast, our information combination reconciled forecast given in equation (\ref{eq:base}) can be written as:
\begin{align*}
	\ytilde_{T+h} & = \bm{\hat{B}}' \yhat_{T+h}  = \yhat_{T+h} + (\bm{\hat{B}}-\bm{I}_m)' \yhat_{T+h} \\
	& =  \yhat_{T+h} + \left(\Yb' \Xb (\Xb'\Xb)^{-1}-\Xb' \Xb (\Xb'\Xb)^{-1}\right) \yhat_{T+h} \\
	& = \yhat_{T+h} +  \left((\Yb -\Xb)' \Xb (\Xb'\Xb)^{-1}\right) \yhat_{T+h} = \yhat_{T+h} +  \left(\Eb' \Xb (\Xb'\Xb)^{-1}\right) \yhat_{T+h}.
\end{align*}
Focusing on the bottom-level series\footnote{Alternatively, with some simple additional algebra, one can write the vector of MinT reconciled forecasts as $\ytilde_{T+h} = \yhat_{T+h} +  \bm{\Psi}^{*'} (\yhat^{top}_{T+h} - \Cb \yhat^{bot}_{T+h})$, where the $i$-th column of $\bm{\Psi}^*$ contains the estimated parameters of the regression of the base forecast error in the $i$-th component of $\yb_t$ on the $m-n$ forecast incoherencies $\yhat_t^{top} - \Cb \yhat_t^{bot}$.}, this leads to
\begin{equation*}
	\ytilde_{T+h}^{bot} = \yhat_{T+h}^{bot} + \bm{\Phi}' \yhat_{T+h},
\end{equation*}
where the $i$-th column of $\bm{\Phi}$ contains the estimated parameters of the regression of the historical forecast errors in the $i$-th component of $\yb_t^{bot}$ on the historical forecasts of \textit{all} $m$ components. This clearly shows that the information combination method allows for the possibility that different components of the base forecasts are based on different information sets. 

We should also highlight that the information combination method (without shrinkage or variable selection) is identical to the approach discussed in \citet{wick2021} as ``EMinTU'' (empirical MinT unconstrained). These methods are closely related to the ERM (empirical risk minimiser) approach introduced by \citet{Ben2019}. ERM uses a holdout validation set to estimate $\Bb$, whereas the former approaches rely on in-sample observations and base forecasts. \citet{Ben2019} suggested using the thin singular value decomposition of $\bm{X}$ to estimate the coefficients when $\bm{X}$ is singular.

The information combination method involves regressions with many more predictors than the MinT sample method, and hence the need for shrinkage or multivariate lasso to obtain better forecasts becomes even more critical. This motivates our proposed alternatives to existing forecast reconciliation methods: the ``information combination with shrinkage'' (ICombS) and ``information combination with lasso'' (ICombL) methods. 

\citet{wang2025} proposed a similar approach, in which they excluded ``poor'' base forecasts (those arising from severe model misspecification or low forecastability of the series) from the reconciliation process, while adjusting the contributions of the remaining series when generating bottom-level reconciled forecasts. They were interested in imposing this sparse structure on the matrix $\bm{G}$, whereas in this paper we focus on the sparsity on $\bm{S}\bm{G}$. \citet{wang2025} formulated this problem as a second-order cone programming problem and solved it using the commercial solver Gurobi. In contrast, our proposed methods can be applied using publicly available software such as the \texttt{glmnet} package.

\section{Proof of concept of information combination using a simple simulation}\label{sec:sim}

In this section, we use a simple example to illustrate the results of the previous section and clarify the role of information sets in forecasting hierarchical time series. This example is designed to expose a source of confusion in the literature that we believe arises from insufficient attention to information sets. Theoretical results in the literature are typically based on the premise that base forecasts are unbiased conditional on a single ensemble information set, as stated in equation \ref{eq:unbiased}. However, simulation studies in the same literature often use base forecasts that are generated from different information sets. In this section, we show that when base forecasts rely on distinct information sets, there is scope to achieve further reductions in the sum of mean squared reconciled forecast errors from an information combination perspective. Parenthetically, by employing a data-generating process in which the top-down reconciliation of unbiased forecasts produces unbiased forecasts, we demonstrate that the often repeated claim that the top-down method can \textit{never} produce unbiased forecasts is incorrect. 

We consider the following data-generating process (DGP) for the hierarchy introduced in Section \ref{sec:relationships}: 
\begin{align*}
	f_t & = 0.6 f_{t-1} + \varepsilon_{1,t} \\
	y_{2,t} & = 1 + 0.8 f_t + \varepsilon_{2,t} \\
	y_{3,t} & = 1 + 0.8 f_t + \varepsilon_{3,t} \\
	y_{1,t} & =  y_{2,t}+y_{3,t}  \\
	\begin{bmatrix}
		\varepsilon_{1,t} \\
		\varepsilon_{2,t} \\
		\varepsilon_{3,t} 
	\end{bmatrix}
	& \sim i.i.d.\ N(\bm{0},\bm{I}_3), \;\;\; \forall t.
\end{align*}
 
In this DGP, $f_t$ is an unobserved component; that is, $\mathcal{I}_T$ contains only the observed history of $\lbrace y_{1,t}, y_{2,t}, y_{3,t} \rbrace$ and not $f_t$ for $t=1,\ldots,T$. Suppose that we have a set of base forecasts, $\yhat_{T+h|T}$, that are conditionally unbiased but not coherent. Then it must be that
\begin{align*}
	E(\hat{y}_{1,T+h|T} \mid \mathcal{I}_T) & = 2 + 1.6 E(f_{T+h} \mid \mathcal{I}_T) \\
	E(\hat{y}_{2,T+h|T} \mid \mathcal{I}_T) & = 1 + 0.8 E(f_{T+h} \mid \mathcal{I}_T) \\
	E(\hat{y}_{3,T+h|T} \mid \mathcal{I}_T) & = 1 + 0.8 E(f_{T+h} \mid \mathcal{I}_T). 
\end{align*}
This implies that a top-down reconciliation with 
\begin{equation*}
	\Gb = \begin{bmatrix}
		0.5 & 0 & 0 \\
		0.5 & 0 & 0 
	\end{bmatrix} \Rightarrow
	\ytilde_{T+h|T} = \Sb \Gb \yhat_{T+h|T} = \begin{bmatrix}
		1 & 0 & 0 \\
		0.5 & 0 & 0 \\
		0.5 & 0 & 0 
	\end{bmatrix} \yhat_{T+h|T}
\end{equation*} 
provides unbiased reconciled forecasts. This example serves as a counterexample to the frequently stated claim that ``the top-down method can \textit{never} produce  unbiased forecasts even if the base forecasts are unbiased'', which appears in many papers on hierarchical forecasting. In Appendix \ref{app:A}, we show that this misconception arises from the incorrect assumption that $\Sb \Gb \Sb = \Sb$ is a necessary (as well as a sufficient) condition for unbiasedness.

Next, we assume that the base forecasts are produced from the implied univariate time series models for each component. This is the scenario considered in most simulation and empirical exercises in the hierarchical forecasting literature. However, to eliminate the effects of the identification methods, estimation algorithms, and sampling variability in parameter estimates, we use the actual conditional expectation function of each series given its own history implied by the data-generating process:
\begin{align*}
y_{1,t} & = 0.8 + 0.6 y_{1,t-1} - 0.24 \eta_{1,t-1} + \eta_{1,t}, \;\; Var(\eta_{1,t})=4.99 \\
y_{2,t} & = 0.4 + 0.6 y_{2,t-1} - 0.33 \eta_{2,t-1} + \eta_{2,t}, \;\; Var(\eta_{2,t})=1.8	\\
y_{3,t} & = 0.4 + 0.6 y_{3,t-1} - 0.33 \eta_{3,t-1} + \eta_{3,t}, \;\; Var(\eta_{3,t})=1.8.	
\end{align*} 
The forecast function for each of these ARMA(1,1) processes is a constant plus a linear combination of its own past values, with exponentially decaying weights as we move further back in time. Specifically, the weight of $y_{i,T-j}$ is $(0.6-\theta)\theta^{j-1}$, where $\theta$ equals $0.24$ for the top series and $0.33$ for the two bottom series. By inspecting these weights, it is clear that the forecasts of the bottom-level series do not add up to the forecast of the top-level series. Note also that these forecasts are not conditionally unbiased in the sense of equation (\ref{eq:unbiased}); hence, they do not satisfy the condition under which MinT is the optimal procedure for minimising the trace of the MSE matrix of the reconciled forecasts. Nevertheless, since MinT does combine forecasts, it may still produce better forecasts than the base forecasts. IComb, on the other hand, does not assume unbiasedness, and obtains the optimal coherent combination of forecasts.

Based on 10,000 one-step-ahead forecasts, the sum of the mean squared forecast error (MSFE) of the base forecasts is $8.600$, which is close to the sum of the variances of the white noise processes in the univariate representations. This is not surprising since we used the actual parameters rather than the estimated ones. The OLS reconciliation reduces the sum of the MSFE to $8.583$. The ideal MinT sample procedure (when the actual $\Wb_1$ implied by the DGP and its associated univariate ARMA forecasts is used in equation \ref{eq:mint}, rather than an estimated one) further reduces this to $8.569$. The ideal IComb procedure (when the information combination regression coefficients implied by the DGP and its associated univariate ARMA forecasts are used) further reduces it to $8.484$. It is important to note that none of these forecasts is conditionally unbiased. Even IComb does not achieve what would be possible if the base forecasts were made conditional on $\mathcal{I}_T$. Although IComb combines information in the three univariate forecasts, each of these forecasts is a particular linear combination of historical values, and therefore the information contained in each is only a subset of $\mathcal{I}_T$. It should be apparent that the IComb strategy also allows users to add other forecasts to the pool. In fact, if a reasonably long history of the hierarchical time series is available, it is plausible that a sophisticated user may wish to generate alternative forecasts as well, though we do not consider that possibility here.

A more practically relevant and interesting scenario arises when the forecast of one component includes information outside of $\mathcal{I}_T$. For example, in the context of forecasting GDP and its components, the Finance Ministry might have information about the schedule of government payments in the next four quarters and incorporate that into its forecast of government expenditure. Similarly, the Central Bank, knowing that interest rate changes take two quarters to affect investment, may produce forecasts of future investment that, in addition to historical data on GDP components, also include recent or upcoming changes in interest rates. In such cases, the objective is not only to produce coherent forecasts but also to allow any forecast containing forward-looking information to improve other forecasts that lack that information. 

To investigate this in our example, we consider the case where $\hat{y}_{1,T+1|T}$ and $\hat{y}_{2,T+1|T}$ are the univariate forecasts as before, but $\hat{y}_{3,T+1|T} = 1+0.8f_{T+1}$. It is clear that in this case, the coefficient of $\hat{y}_{3,T+1|T}$ in the IComb regressions will be highly significant. In fact, theoretically, the other two forecasts should receive zero weight. In 10,000 simulations, the sums of the MSFEs are $7.776$ for the base forecasts, $7.508$ for the OLS-reconciled forecasts, $4.343$ for MinT sample, and $3.952$ for IComb. The minimum achievable sum of one-step-ahead MSFEs in this DGP, when $f_T$ is known, is $4$, and IComb attains that value (with a small discrepancy due to simulation error). Note that while MinT substantially improves the forecasts, it relies only on the information contained in the incoherency term $y_{1,t}-y_{2,t}-y_{3,t}$, not on $y_{3,t}$ itself, and therefore cannot reach the minimum that IComb achieves.

In this simple example, we have deliberately avoided parameter estimation in order to obtain a focused evaluation of the role of different information sets and to distinguish between the value of information combination and that of forecast reconciliation. We defer the investigation of how the algorithms that generate univariate forecasts, the sample size, the dimensionality of the time series, and the use of shrinkage affect the results. These issues are discussed in the next section.

\section{Empirical investigation}
\label{sec:empirical}
In this section, we investigate the performance of the information combination methods using Australian domestic tourism data, which have been used in several studies, including \citet{HyndmanEtAl2011} and \citet{WicEtAl2019}.

Domestic tourism flow is measured using \textit{visitor nights}, which represent the total number of nights Australians spend away from home. These data are maintained by Tourism Research Australia and are sourced from the National Visitor Survey, which is conducted via computer-assisted telephone interviews. Each year, the survey collects information from a sample of 120,000 Australian residents aged 15 and over. The data used in this study form a monthly time series covering the period from January 1998 to December 2019.

Using this information, we construct a hierarchy with three levels. The total visitor nights in Australia is first disaggregated by seven states and territories: New South Wales (NSW), Victoria (VIC), Queensland (QLD), South Australia (SA),
Western Australia (WA), Tasmania (TAS), and the Northern Territory (NT). Each of these is then disaggregated by zones (27 series in total) and then by regions (77 series in total). The hierarchy therefore contains 112 series in total.

For each series in the hierarchy, we begin with a training set of size 120 and fit univariate autoregressive integrated moving average (ARIMA) and \{error, trend, seasonal\} (ETS) models by minimising the corrected AIC (AICc). We use the default settings implemented in the \texttt{forecast} package \citep{hyndman2024}.
The base forecasts are then produced for 1- to 12-steps-ahead for each series in the hierarchy. The base forecasts are reconciled using several alternative approaches, namely, the bottom-up (BU), ordinary least squares (OLS), weighted least squares with variance scaling ($\text{WLS}_v$) \citep{hyndman2016fast}, MinT using the shrinkage covariance estimate, and EMinTU methods. 

For the IComb methods, the optimal tuning parameter must be determined, which we achieve through a grid search procedure. We construct a grid of size 200 that is linear on the log scale and set the maximum value of the tuning parameter (denoted $\tau_{max}$) using the closed-form solution as in the \texttt{glmnet} package. The minimum value of the tuning parameter is set to $0.01\,\tau_{max} \times 10^{-\lfloor\log_{10}(\tau_{max})\rfloor}$, where $\lfloor \cdot \rfloor$ denotes the floor operator of a real-valued number, whereas \texttt{glmnet} always sets it to $0.0001\,\tau_{max}$. For each value in the grid, we divide the training set into another training set (we refer to this as the \textit{second training set} in the following discussion) and a validation set. The validation set consists of the last 40 observations of the initial training set. Similarly, we divide the 1-step-ahead forecasts in the training set into two parts. Using the observations and the corresponding 1-step-ahead forecasts in the second training set, we compute the IComb coefficients using \texttt{glmnet}. The estimated coefficients are then used to reconcile the first 1-step-ahead forecast in the validation set, and the reconciled forecast error is calculated. We roll the second training set forward by one observation and repeat the steps described above until the end of the validation set is reached. Then, we compute the mean squared reconciled forecast error over the validation set, and the tuning parameter that minimises this mean squared error is selected as the optimal tuning parameter. Finally, using the selected tuning parameter, together with the observations in the initial training set and the corresponding 1-step-ahead forecasts, we estimate the IComb coefficients with  \texttt{glmnet}, and these coefficients are then used to calculate the IComb reconciled forecasts.

We roll the training window forward by one observation at a time until November 2019. After reconciling the base forecasts using IComb and the alternative methods, we evaluate their accuracy using the mean squared forecast error (MSFE). To measure the improvement in forecast accuracy, we compute the percentage relative improvement in average loss (PRIAL) for each method relative to the base forecasts, defined as follows:
\[
\text{PRIAL}_{\text{method}} = \frac{\text{MSFE}_{\text{method}} - \text{MSFE}_{\text{base}}}{\text{MSFE}_{\text{base}}}\times 100.
\]

These results are summarised in Tables \ref{tbl:prial-arima} and \ref{tbl:prial-ets}, corresponding to the base forecasts produced by the ARIMA and ETS models, respectively. Both tables are identically structured, reporting the PRIAL values for each reconciliation method across different hierarchical levels and forecast horizons from 1- to 12-steps-ahead. We also report the overall performance for the short-term ($h=1$--6) and long-term ($h=1$--12) forecast horizons, where PRIAL values are computed from the averaged MSFEs of the corresponding forecast steps. Each panel in these tables corresponds to a specific level of the hierarchy, beginning with the top level denoted as ``Australia'', followed by disaggregated levels such as State, Zones, and Regions. The last panel with the top row labelled ``Average'', reports the PRIALs computed from the MSFEs averaged across all series in the hierarchy. Within each panel, the row labelled ``Base'' reports the MSFEs of the base forecasts, with the actual values scaled by a factor of $10^4$. The rows above display the PRIAL values for each reconciliation method relative to the base forecasts. A positive (negative) value denotes a percentage increase (decrease) in the MSFE relative to the base forecasts. The bold entries highlight the largest negative PRIAL values, signifying the best-performing reconciliation method for the corresponding panel.

The IComb method consists of several variants, defined based on three factors: (i) the penalty ($\alpha = 1$ for multivariate lasso and $\alpha = 0$ for ridge regression), (ii) the type of standardisation applied ($z_{x,y} = 1$ if both $\bm{X}$ and $\bm{Y}$ are standardised, $z_{x} = 1$ if only $\bm{X}$ is standardised, and $z_{x,y} = 0$ if neither is standardised), and (iii) the inclusion of the intercept ($c = 1$ if included and $c = 0$ otherwise). The results for $c = 0$ are provided in Appendix \ref{app:all-results}. The impact of these IComb variants on forecast accuracy is discussed in detail in Sections \ref{sec:arima} and \ref{sec:ets}, which present the out-of-sample forecast performance for ARIMA and ETS forecasts, respectively.

\subsection{ARIMA forecasts}
\label{sec:arima}


Table \ref{tbl:prial-arima} reports the PRIAL for various reconciliation methods applied to ARIMA base forecasts. The results demonstrate that the IComb method outperforms other reconciliation approaches, as reflected in more negative PRIAL values across most hierarchical levels and forecast horizons. Although some IComb variants fail to improve the base forecasts in a few cases, the overall pattern still shows significant improvements in the forecasts reconciled by IComb. Notably, the variants with ridge regression consistently outperform those with multivariate lasso regression, both at the individual levels and when averaged across all hierarchical levels. 

In particular, the IComb variant using ridge regression with an intercept and unstandardised predictor and response variables achieves the most substantial gains in forecast accuracy, showing the most negative PRIAL across most forecast horizons. At the national level, while other reconciliation approaches apart from IComb hardly improve forecast accuracy, this IComb variant demonstrates consistent performance advantages. Although it is slightly outperformed by other IComb variants at forecast horizons 2, 7, 11, and 12, it generally achieves the greatest reduction in forecast errors among all reconciliation methods, as evidenced by the PRIAL values. Its largest improvement occurs at the forecast horizon of 8, where the forecast error of the base forecasts is reduced by 10.28$\%$. This indicates the greatest gain in forecast accuracy among all reconciliation techniques and forecast horizons at this hierarchical level. This IComb variant also secures the best average performance over both short-term ($h=1$--$6$) and long-term ($h=1$--$12$), with PRIAL values of $-5.36$ and $-6.34$, respectively. This strong performance persists across disaggregated levels and in the hierarchical average, where the variant consistently achieves improvements of more than 10$\%$ relative to the base forecasts.

The enhanced performance of ridge regression over multivariate lasso in this context suggests that each base forecast contains some useful information that can improve the forecasts of other components. Therefore, a dense model, such as that produced by ridge regression, yields better forecasts than a sparse model, such as that produced by lasso. There are also substantial similarities among the bottom-level series within the same region, as tourists commonly visit multiple destinations in one region during a single trip. This results in a high degree of collinearity among the ARIMA forecasts of these series. It is well known that lasso performs poorly when predictors are highly collinear.

The differences in parsimony among lasso-based IComb variants under alternative standardisation types and intercept specifications are explored in the top panel of Figure \ref{fig:nonzeros}. This panel displays the kernel density estimates of the number of non-zero rows in the estimated coefficient matrices across IComb variants using multivariate lasso regression. A rightward shift in the distribution is noticeable as standardisation is applied to more variables, indicating that more regressors are retained. In each case, the exclusion of the intercept results in a further slight rightward shift of the distribution, indicating the inclusion of more regressors in the model. Interestingly, as the lasso-based models become less parsimonious, their PRIAL results do not approach those of their ridge-based counterparts. This may be because, in lasso regularisation, a single tuning parameter controls both parsimony and shrinkage. The lower degree of shrinkage in less parsimonious models leads to worse performance in this context, where shrinkage appears to be more important than parsimony.

\begin{table}[h]
\caption{PRIAL for forecast reconciliation methods using ARIMA base forecasts.}
\label{tbl:prial-arima}
\centering
\resizebox{\ifdim\width>\linewidth\linewidth\else\width\fi}{!}{
\begin{tabular}[t]{lrrrrrrrrrrrrrr}
\toprule
& \multicolumn{14}{c}{Forecast horizon $(h)$}\\
\cmidrule{2-15}
  & {\it 1} & {\it 2} & {\it 3} & {\it 4} & {\it 5} & {\it 6} & {\it 7} & {\it 8} & {\it 9} & {\it 10} & {\it 11} & {\it 12} & {\it 1--6} & {\it 1--12}\\
\cmidrule{2-15}
\multicolumn{15}{l}{\it Australia}\\
\hspace{1em}BU & $53.73$ & $62.64$ & $63.58$ & $52.97$ & $56.61$ & $56.62$ & $62.85$ & $59.57$ & $72.53$ & $60.50$ & $52.84$ & $55.91$ & $57.56$ & $59.06$\\
\hspace{1em}OLS & $0.01$ & $-0.01$ & $0.68$ & $0.02$ & $0.22$ & $-0.05$ & $1.03$ & $0.38$ & $2.37$ & $1.40$ & $0.57$ & $1.16$ & $0.14$ & $0.68$\\
\hspace{1em}WLS$_v$ & $23.46$ & $27.45$ & $31.40$ & $23.51$ & $26.06$ & $25.25$ & $31.01$ & $28.20$ & $38.62$ & $30.26$ & $24.63$ & $27.85$ & $26.12$ & $28.16$\\
\hspace{1em}MinT & $13.86$ & $14.48$ & $20.16$ & $13.27$ & $14.22$ & $12.73$ & $18.68$ & $15.21$ & $24.01$ & $17.82$ & $12.91$ & $16.50$ & $14.71$ & $16.14$\\
\hspace{1em}EMinTU & $362.12$ & $424.60$ & $406.27$ & $324.13$ & $280.36$ & $300.57$ & $304.59$ & $300.86$ & $281.66$ & $285.19$ & $152.46$ & $262.24$ & $346.43$ & $300.35$\\
\hspace{1em}IComb$[\alpha = 1, z_{x, y} = 1, c = 1]$ & $5.73$ & $8.44$ & $15.71$ & $4.67$ & $6.38$ & $5.65$ & $10.02$ & $1.91$ & $13.86$ & $7.31$ & $-1.86$ & $3.18$ & $7.64$ & $6.44$\\
\hspace{1em}IComb$[\alpha = 1, z_{x} = 1, c = 1]$ & $-2.42$ & $-1.45$ & $12.48$ & $1.40$ & $2.04$ & $-2.85$ & $-1.59$ & $-5.53$ & $5.12$ & $0.97$ & $-5.34$ & $1.29$ & $1.47$ & $0.17$\\
\hspace{1em}IComb$[\alpha = 1, z_{x, y} = 0, c = 1]$ & $-0.90$ & $\pmb{-7.64}$ & $1.72$ & $-4.22$ & $-4.73$ & $-7.79$ & $\pmb{-7.46}$ & $-7.47$ & $-0.33$ & $-3.70$ & $-6.67$ & $-2.14$ & $-4.04$ & $-4.36$\\
\hspace{1em}IComb$[\alpha = 0, z_{x, y} = 1, c = 1]$ & $2.80$ & $5.48$ & $10.09$ & $3.32$ & $3.99$ & $0.94$ & $0.59$ & $-2.65$ & $4.98$ & $-1.80$ & $\pmb{-9.68}$ & $\pmb{-5.20}$ & $4.33$ & $0.56$\\
\hspace{1em}IComb$[\alpha = 0, z_{x} = 1, c = 1]$ & $2.96$ & $5.69$ & $10.14$ & $3.35$ & $3.97$ & $0.99$ & $0.48$ & $-2.66$ & $4.87$ & $-1.99$ & $-9.61$ & $-5.15$ & $4.41$ & $0.57$\\
\hspace{1em}IComb$[\alpha = 0, z_{x, y} = 0, c = 1]$ & $\pmb{-3.17}$ & $-5.27$ & $\pmb{-1.08}$ & $\pmb{-5.46}$ & $\pmb{-6.95}$ & $\pmb{-9.35}$ & $-6.96$ & $\pmb{-10.28}$ & $\pmb{-5.67}$ & $\pmb{-6.66}$ & $-8.76$ & $-4.70$ & $\pmb{-5.36}$ & $\pmb{-6.34}$\\
\hspace{1em}Base $(\times 10^4)$ & {\it 365.04} & {\it 367.90} & {\it 378.85} & {\it 422.84} & {\it 419.69} & {\it 436.30} & {\it 437.85} & {\it 459.88} & {\it 435.16} & {\it 486.94} & {\it 519.41} & {\it 522.46} & {\it 398.44} & {\it 437.69}\\
\cmidrule{2-15}
\multicolumn{15}{l}{\it States}\\
\hspace{1em}BU & $21.15$ & $22.50$ & $22.96$ & $23.16$ & $24.97$ & $24.49$ & $23.57$ & $24.36$ & $23.08$ & $21.39$ & $20.21$ & $19.11$ & $23.24$ & $22.50$\\
\hspace{1em}OLS & $-4.38$ & $-4.42$ & $-5.67$ & $-3.25$ & $-3.80$ & $-3.24$ & $-4.80$ & $-3.91$ & $-6.25$ & $-5.13$ & $-4.00$ & $-4.94$ & $-4.11$ & $-4.50$\\
\hspace{1em}WLS$_v$ & $4.45$ & $4.97$ & $5.54$ & $6.53$ & $7.21$ & $7.24$ & $6.61$ & $6.82$ & $5.99$ & $5.36$ & $5.25$ & $4.90$ & $6.02$ & $5.90$\\
\hspace{1em}MinT & $0.22$ & $-0.83$ & $0.39$ & $1.72$ & $1.62$ & $1.41$ & $0.94$ & $0.67$ & $-0.27$ & $-0.18$ & $-0.20$ & $-0.20$ & $0.77$ & $0.42$\\
\hspace{1em}EMinTU & $541.01$ & $474.48$ & $448.10$ & $392.24$ & $335.27$ & $353.31$ & $350.76$ & $322.01$ & $304.31$ & $295.59$ & $271.75$ & $294.01$ & $422.14$ & $359.60$\\
\hspace{1em}IComb$[\alpha = 1, z_{x, y} = 1, c = 1]$ & $-4.95$ & $-3.80$ & $-0.63$ & $-1.91$ & $-2.51$ & $-1.52$ & $-2.78$ & $-4.74$ & $-3.71$ & $-3.93$ & $-6.02$ & $-5.14$ & $-2.53$ & $-3.54$\\
\hspace{1em}IComb$[\alpha = 1, z_{x} = 1, c = 1]$ & $-6.78$ & $-6.48$ & $-2.16$ & $-2.02$ & $-4.64$ & $-5.39$ & $-8.23$ & $-8.22$ & $-7.23$ & $-6.88$ & $-7.58$ & $-6.23$ & $-4.56$ & $-6.07$\\
\hspace{1em}IComb$[\alpha = 1, z_{x, y} = 0, c = 1]$ & $-8.09$ & $-11.12$ & $-8.88$ & $-7.56$ & $-8.65$ & $-9.30$ & $-12.33$ & $-11.53$ & $-11.44$ & $-11.69$ & $-10.47$ & $-9.99$ & $-8.93$ & $-10.15$\\
\hspace{1em}IComb$[\alpha = 0, z_{x, y} = 1, c = 1]$ & $-6.49$ & $-4.58$ & $-3.58$ & $-3.73$ & $-3.60$ & $-4.02$ & $-7.22$ & $-7.38$ & $-8.89$ & $-10.04$ & $-11.47$ & $-10.55$ & $-4.32$ & $-7.00$\\
\hspace{1em}IComb$[\alpha = 0, z_{x} = 1, c = 1]$ & $-6.35$ & $-4.48$ & $-3.57$ & $-3.68$ & $-3.59$ & $-4.03$ & $-7.30$ & $-7.36$ & $-8.98$ & $-10.10$ & $-11.43$ & $-10.52$ & $-4.26$ & $-6.99$\\
\hspace{1em}IComb$[\alpha = 0, z_{x, y} = 0, c = 1]$ & $\pmb{-9.42}$ & $\pmb{-11.25}$ & $\pmb{-10.40}$ & $\pmb{-9.02}$ & $\pmb{-10.22}$ & $\pmb{-10.38}$ & $\pmb{-12.63}$ & $\pmb{-13.13}$ & $\pmb{-14.19}$ & $\pmb{-13.14}$ & $\pmb{-11.82}$ & $\pmb{-11.29}$ & $\pmb{-10.11}$ & $\pmb{-11.48}$\\
\hspace{1em}Base $(\times 10^4)$ & {\it 27.44} & {\it 27.80} & {\it 28.19} & {\it 28.86} & {\it 29.04} & {\it 29.96} & {\it 31.03} & {\it 31.37} & {\it 32.22} & {\it 33.56} & {\it 34.36} & {\it 35.23} & {\it 28.55} & {\it 30.75}\\
\cmidrule{2-15}
\multicolumn{15}{l}{\it Zones}\\
\hspace{1em}BU & $4.33$ & $4.49$ & $3.45$ & $3.79$ & $4.77$ & $3.51$ & $4.67$ & $5.34$ & $4.55$ & $4.57$ & $2.23$ & $2.21$ & $4.05$ & $3.98$\\
\hspace{1em}OLS & $-6.32$ & $-6.51$ & $-7.73$ & $-6.68$ & $-6.93$ & $-7.24$ & $-6.84$ & $-6.34$ & $-7.34$ & $-6.63$ & $-6.45$ & $-6.66$ & $-6.91$ & $-6.80$\\
\hspace{1em}WLS$_v$ & $-3.80$ & $-4.12$ & $-5.09$ & $-4.53$ & $-4.32$ & $-4.76$ & $-3.89$ & $-3.67$ & $-4.07$ & $-3.77$ & $-4.53$ & $-4.37$ & $-4.44$ & $-4.24$\\
\hspace{1em}MinT & $-6.41$ & $-7.09$ & $-8.00$ & $-7.26$ & $-7.25$ & $-7.77$ & $-6.86$ & $-6.77$ & $-7.21$ & $-6.70$ & $-7.49$ & $-7.10$ & $-7.30$ & $-7.16$\\
\hspace{1em}EMinTU & $539.68$ & $461.84$ & $454.25$ & $391.82$ & $356.57$ & $354.69$ & $356.60$ & $344.48$ & $336.82$ & $325.06$ & $345.24$ & $317.08$ & $425.20$ & $380.05$\\
\hspace{1em}IComb$[\alpha = 1, z_{x, y} = 1, c = 1]$ & $-12.08$ & $-11.77$ & $-11.25$ & $-12.12$ & $-11.95$ & $-11.95$ & $-10.57$ & $-10.72$ & $-10.54$ & $-9.56$ & $-10.99$ & $-9.86$ & $-11.85$ & $-11.09$\\
\hspace{1em}IComb$[\alpha = 1, z_{x} = 1, c = 1]$ & $-12.18$ & $-12.12$ & $-11.36$ & $-11.03$ & $-11.94$ & $-12.51$ & $-12.30$ & $-11.44$ & $-11.28$ & $-10.24$ & $-10.78$ & $-9.57$ & $-11.86$ & $-11.38$\\
\hspace{1em}IComb$[\alpha = 1, z_{x, y} = 0, c = 1]$ & $-10.86$ & $-13.52$ & $-13.06$ & $-11.99$ & $-12.93$ & $-12.91$ & $-13.57$ & $-12.71$ & $-12.64$ & $-11.75$ & $-11.30$ & $-10.02$ & $-12.55$ & $-12.25$\\
\hspace{1em}IComb$[\alpha = 0, z_{x, y} = 1, c = 1]$ & $-12.79$ & $-12.63$ & $-12.64$ & $-13.59$ & $-13.09$ & $-13.66$ & $-13.24$ & $-12.53$ & $-14.00$ & $-13.77$ & $\pmb{-14.99}$ & $\pmb{-13.75}$ & $-13.07$ & $-13.41$\\
\hspace{1em}IComb$[\alpha = 0, z_{x} = 1, c = 1]$ & $-12.74$ & $-12.59$ & $-12.62$ & $-13.56$ & $-13.10$ & $-13.67$ & $-13.30$ & $-12.52$ & $-14.05$ & $-13.79$ & $\pmb{-14.99}$ & $\pmb{-13.75}$ & $-13.05$ & $-13.41$\\
\hspace{1em}IComb$[\alpha = 0, z_{x, y} = 0, c = 1]$ & $\pmb{-14.28}$ & $\pmb{-15.95}$ & $\pmb{-15.91}$ & $\pmb{-15.30}$ & $\pmb{-15.52}$ & $\pmb{-15.44}$ & $\pmb{-14.97}$ & $\pmb{-15.51}$ & $\pmb{-15.97}$ & $\pmb{-14.49}$ & $-14.62$ & $-13.03$ & $\pmb{-15.40}$ & $\pmb{-15.07}$\\
\hspace{1em}Base $(\times 10^4)$ & {\it 4.99} & {\it 5.05} & {\it 5.14} & {\it 5.21} & {\it 5.21} & {\it 5.31} & {\it 5.32} & {\it 5.33} & {\it 5.40} & {\it 5.47} & {\it 5.57} & {\it 5.61} & {\it 5.15} & {\it 5.30}\\
\cmidrule{2-15}
\multicolumn{15}{l}{\it Regions}\\
\hspace{1em}BU & $0.00$ & $0.00$ & $0.00$ & $0.00$ & $0.00$ & $0.00$ & $0.00$ & $0.00$ & $0.00$ & $0.00$ & $0.00$ & $0.00$ & $0.00$ & $0.00$\\
\hspace{1em}OLS & $-4.32$ & $-4.84$ & $-5.13$ & $-4.71$ & $-5.16$ & $-4.73$ & $-4.99$ & $-5.08$ & $-5.32$ & $-4.88$ & $-3.81$ & $-4.07$ & $-4.82$ & $-4.75$\\
\hspace{1em}WLS$_v$ & $-3.49$ & $-4.03$ & $-4.13$ & $-4.00$ & $-4.35$ & $-4.12$ & $-4.13$ & $-4.52$ & $-4.24$ & $-4.10$ & $-3.47$ & $-3.36$ & $-4.02$ & $-4.00$\\
\hspace{1em}MinT & $-4.74$ & $-5.73$ & $-5.93$ & $-5.84$ & $-6.19$ & $-5.89$ & $-5.90$ & $-6.47$ & $-6.04$ & $-5.91$ & $-5.20$ & $-4.95$ & $-5.72$ & $-5.73$\\
\hspace{1em}EMinTU & $560.81$ & $512.11$ & $462.76$ & $412.31$ & $396.78$ & $380.79$ & $375.80$ & $377.99$ & $364.11$ & $363.97$ & $386.29$ & $359.63$ & $453.55$ & $411.47$\\
\hspace{1em}IComb$[\alpha = 1, z_{x, y} = 1, c = 1]$ & $-12.94$ & $-13.36$ & $-12.60$ & $-13.06$ & $-13.28$ & $-12.85$ & $-12.40$ & $-12.48$ & $-12.34$ & $-10.95$ & $-11.08$ & $-10.40$ & $-13.02$ & $-12.29$\\
\hspace{1em}IComb$[\alpha = 1, z_{x} = 1, c = 1]$ & $-13.27$ & $-13.85$ & $-13.01$ & $-12.62$ & $-12.97$ & $-13.06$ & $-13.05$ & $-12.57$ & $-12.52$ & $-11.16$ & $-10.39$ & $-10.07$ & $-13.13$ & $-12.36$\\
\hspace{1em}IComb$[\alpha = 1, z_{x, y} = 0, c = 1]$ & $-10.47$ & $-12.83$ & $-12.19$ & $-11.86$ & $-12.26$ & $-11.86$ & $-12.38$ & $-11.98$ & $-11.39$ & $-10.06$ & $-8.67$ & $-7.67$ & $-11.91$ & $-11.11$\\
\hspace{1em}IComb$[\alpha = 0, z_{x, y} = 1, c = 1]$ & $-13.71$ & $-14.02$ & $-13.59$ & $-14.21$ & $-14.12$ & $-14.27$ & $-14.08$ & $-13.95$ & $-14.67$ & $\pmb{-13.43}$ & $\pmb{-13.96}$ & $\pmb{-13.11}$ & $-13.99$ & $-13.92$\\
\hspace{1em}IComb$[\alpha = 0, z_{x} = 1, c = 1]$ & $-13.67$ & $-13.99$ & $-13.57$ & $-14.19$ & $-14.10$ & $-14.27$ & $-14.12$ & $-13.96$ & $-14.70$ & $-13.41$ & $-13.95$ & $\pmb{-13.11}$ & $-13.97$ & $-13.92$\\
\hspace{1em}IComb$[\alpha = 0, z_{x, y} = 0, c = 1]$ & $\pmb{-14.09}$ & $\pmb{-15.24}$ & $\pmb{-15.10}$ & $\pmb{-15.21}$ & $\pmb{-15.05}$ & $\pmb{-14.78}$ & $\pmb{-14.38}$ & $\pmb{-14.71}$ & $\pmb{-14.92}$ & $-13.19$ & $-12.21$ & $-11.10$ & $\pmb{-14.91}$ & $\pmb{-14.14}$\\
\hspace{1em}Base $(\times 10^4)$ & {\it 1.49} & {\it 1.48} & {\it 1.49} & {\it 1.51} & {\it 1.52} & {\it 1.53} & {\it 1.55} & {\it 1.55} & {\it 1.57} & {\it 1.58} & {\it 1.58} & {\it 1.59} & {\it 1.50} & {\it 1.54}\\
\cmidrule{2-15}
\multicolumn{15}{l}{\it Average}\\
\hspace{1em}BU & $30.09$ & $34.51$ & $35.05$ & $31.32$ & $33.52$ & $33.44$ & $36.28$ & $35.54$ & $40.35$ & $35.47$ & $31.63$ & $32.84$ & $32.99$ & $34.19$\\
\hspace{1em}OLS & $-2.71$ & $-2.83$ & $-3.04$ & $-2.42$ & $-2.56$ & $-2.53$ & $-2.37$ & $-2.33$ & $-2.26$ & $-2.12$ & $-2.04$ & $-2.03$ & $-2.68$ & $-2.41$\\
\hspace{1em}WLS$_v$ & $10.55$ & $12.36$ & $14.22$ & $11.52$ & $12.82$ & $12.53$ & $15.21$ & $14.21$ & $18.38$ & $15.07$ & $12.55$ & $14.07$ & $12.34$ & $13.68$\\
\hspace{1em}MinT & $4.58$ & $4.37$ & $7.13$ & $4.83$ & $5.17$ & $4.46$ & $7.29$ & $5.72$ & $9.28$ & $6.99$ & $4.75$ & $6.60$ & $5.08$ & $5.96$\\
\hspace{1em}EMinTU & $462.60$ & $455.04$ & $432.07$ & $362.15$ & $320.65$ & $331.74$ & $332.91$ & $322.22$ & $306.60$ & $303.26$ & $235.94$ & $289.18$ & $391.53$ & $341.15$\\
\hspace{1em}IComb$[\alpha = 1, z_{x, y} = 1, c = 1]$ & $-2.44$ & $-0.93$ & $3.39$ & $-1.85$ & $-1.21$ & $-1.19$ & $0.85$ & $-3.39$ & $2.33$ & $-0.11$ & $-5.25$ & $-2.28$ & $-0.72$ & $-1.08$\\
\hspace{1em}IComb$[\alpha = 1, z_{x} = 1, c = 1]$ & $-6.63$ & $-6.18$ & $1.48$ & $-3.21$ & $-3.72$ & $-6.29$ & $-6.33$ & $-7.95$ & $-2.76$ & $-4.05$ & $-7.25$ & $-3.40$ & $-4.09$ & $-4.73$\\
\hspace{1em}IComb$[\alpha = 1, z_{x, y} = 0, c = 1]$ & $-5.63$ & $\pmb{-10.18}$ & $-5.20$ & $-7.23$ & $-7.95$ & $-9.47$ & $-10.21$ & $-9.79$ & $-6.41$ & $-7.57$ & $-8.46$ & $-5.79$ & $-7.64$ & $-7.83$\\
\hspace{1em}IComb$[\alpha = 0, z_{x, y} = 1, c = 1]$ & $-4.36$ & $-2.70$ & $-0.24$ & $-3.30$ & $-2.89$ & $-4.49$ & $-5.33$ & $-6.69$ & $-3.94$ & $-6.96$ & $\pmb{-11.38}$ & $\pmb{-8.64}$ & $-3.02$ & $-5.28$\\
\hspace{1em}IComb$[\alpha = 0, z_{x} = 1, c = 1]$ & $-4.24$ & $-2.57$ & $-0.21$ & $-3.27$ & $-2.90$ & $-4.47$ & $-5.41$ & $-6.69$ & $-4.03$ & $-7.07$ & $-11.33$ & $-8.60$ & $-2.96$ & $-5.27$\\
\hspace{1em}IComb$[\alpha = 0, z_{x, y} = 0, c = 1]$ & $\pmb{-8.06}$ & $-9.89$ & $\pmb{-7.72}$ & $\pmb{-9.13}$ & $\pmb{-10.15}$ & $\pmb{-11.26}$ & $\pmb{-10.52}$ & $\pmb{-12.30}$ & $\pmb{-10.57}$ & $\pmb{-10.16}$ & $-10.73$ & $-8.22$ & $\pmb{-9.41}$ & $\pmb{-9.93}$\\
\hspace{1em}Base $(\times 10^4)$ & {\it 7.20} & {\it 7.25} & {\it 7.41} & {\it 7.87} & {\it 7.86} & {\it 8.10} & {\it 8.20} & {\it 8.42} & {\it 8.28} & {\it 8.85} & {\it 9.22} & {\it 9.32} & {\it 7.62} & {\it 8.16}\\
\bottomrule
\end{tabular}}
\end{table}

\subsection{ETS forecasts}
\label{sec:ets}


Table \ref{tbl:prial-ets} reports PRIAL for various reconciliation methods applied to the ETS base forecasts. The results show significantly lower MSFEs for ETS base forecasts compared to ARIMA base forecasts, suggesting the presence of trends or seasonal structures in the tourism dataset that the ETS model handles more effectively than the ARIMA model. More importantly, all IComb variants consistently improve the ETS base forecasts, as demonstrated by their negative PRIAL values in Table \ref{tbl:prial-ets}.

The PRIAL values in Table \ref{tbl:prial-ets} reveal that while all IComb methods outperform the other reconciliation methods considered, no single IComb variant consistently outperforms all others. This contrasts with the ARIMA case, where the ridge variant without standardisation performed best in nearly all instances. For the ETS based forecasts, at the national level, the IComb variants using multivariate lasso with standardisation outperform ridge-based IComb variants. Specifically, the lasso-based variant with an intercept and standardised predictors and response variables performs best at the first four forecast horizons, whereas the one with an intercept and standardised predictors performs better at longer horizons. In contrast, the bottom-level forecasts benefit more from the ridge-based IComb variants, which reduce the mean squared forecast errors by more than 5$\%$ in most cases, although the lasso variant with full standardisation is not far behind. 

\begin{figure}[h]
    \centering
    \includegraphics[width=\textwidth]{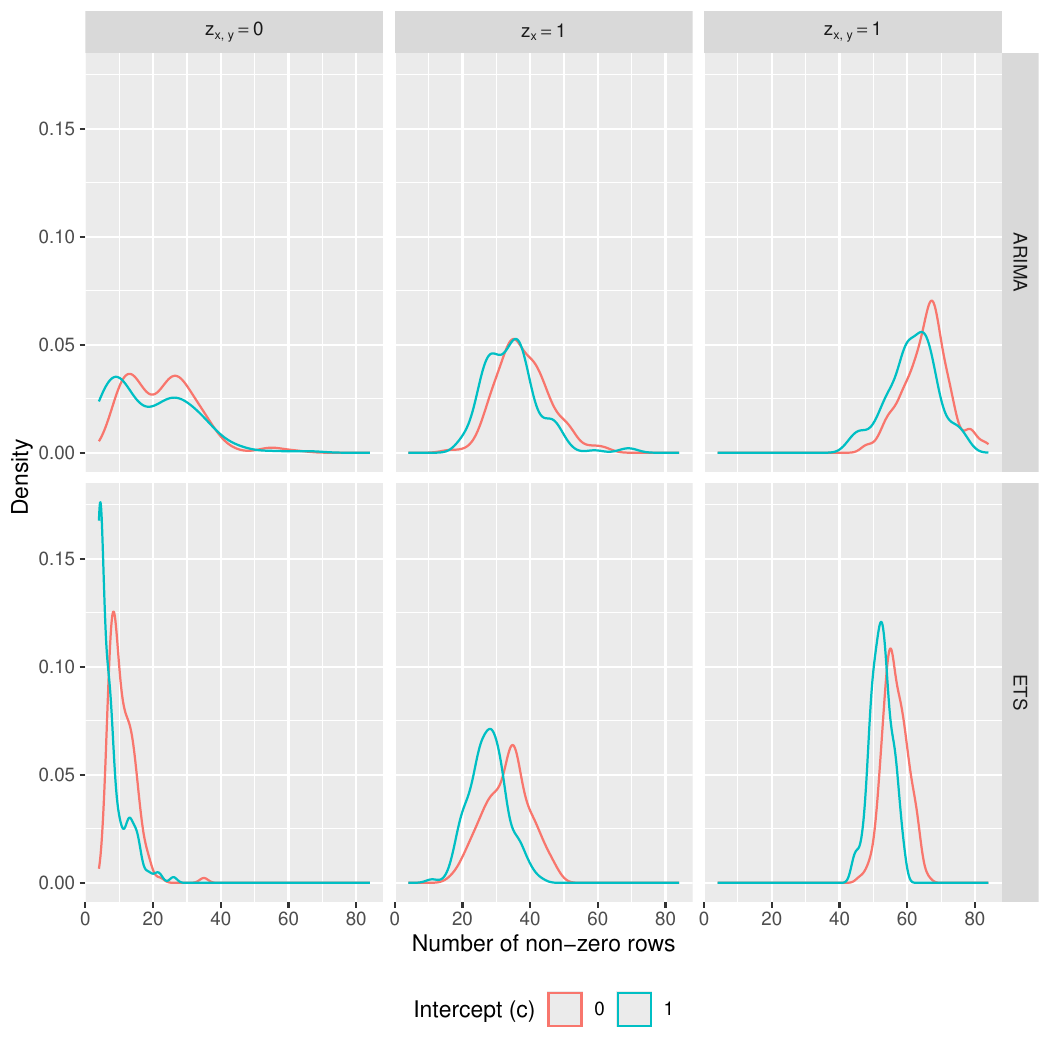}
    \caption{Kernal density estimates of number of non-zero rows in the estimated coefficient matrices for multivariate lasso.}
    \label{fig:nonzeros}
\end{figure}

The degree of parsimony of lasso-based IComb variants under alternative standardisation types and intercept specifications is explored in the bottom panel of Figure \ref{fig:nonzeros}. Similar to the ARIMA case, applying standardisation to more variables shifts the distributions of number of selected predictors to the right, and in each case, the exclusion of the intercept pushes the density slightly further right. Compared with the corresponding densities in the ARIMA case, the densities of the number of predictors chosen in the ETS case are tighter and more peaked, suggesting a more stable set of predictors over time. For example, when both the response variable and the predictors are standardised and an intercept is included, the number of selected predictors for the ETS-based forecasts ranges between 40 and 60, whereas the corresponding range for the ARIMA-based forecasts is 40 to 80. This, together with the PRIAL values, suggests less redundancy and multicollinearity in the ETS base forecasts relative to the ARIMA base forecasts.

\begin{table}[h]
\centering
\caption{PRIAL for forecast reconciliation methods using ETS base forecasts.}
\label{tbl:prial-ets}
\centering
\resizebox{\ifdim\width>\linewidth\linewidth\else\width\fi}{!}{
\begin{tabular}[t]{lrrrrrrrrrrrrrr}
\toprule
& \multicolumn{14}{c}{Forecast horizon $(h)$}\\
\cmidrule{2-15}
  & {\it 1} & {\it 2} & {\it 3} & {\it 4} & {\it 5} & {\it 6} & {\it 7} & {\it 8} & {\it 9} & {\it 10} & {\it 11} & {\it 12} & {\it 1--6} & {\it 1--12}\\
\cmidrule{2-15}
\multicolumn{15}{l}{\it Australia}\\
\hspace{1em}BU & $37.41$ & $28.81$ & $27.54$ & $26.57$ & $29.33$ & $32.47$ & $32.88$ & $39.52$ & $38.26$ & $37.70$ & $38.09$ & $38.99$ & $30.27$ & $34.41$\\
\hspace{1em}OLS & $0.07$ & $-0.31$ & $-0.94$ & $-1.31$ & $-1.29$ & $-0.77$ & $-0.32$ & $0.11$ & $-0.26$ & $-0.23$ & $-0.08$ & $-0.15$ & $-0.78$ & $-0.43$\\
\hspace{1em}WLS$_v$ & $17.65$ & $13.14$ & $11.32$ & $9.59$ & $10.93$ & $13.87$ & $15.76$ & $19.35$ & $18.24$ & $18.00$ & $18.45$ & $18.76$ & $12.66$ & $15.73$\\
\hspace{1em}MinT & $12.91$ & $8.79$ & $6.90$ & $4.67$ & $6.25$ & $8.65$ & $10.55$ & $13.30$ & $12.06$ & $12.19$ & $13.01$ & $12.54$ & $7.93$ & $10.38$\\
\hspace{1em}EMinTU & $792.26$ & $1174.87$ & $1648.94$ & $1785.04$ & $2195.41$ & $2379.61$ & $2909.64$ & $2936.09$ & $2825.55$ & $3066.33$ & $2922.22$ & $3074.02$ & $1693.24$ & $2403.10$\\
\hspace{1em}IComb$[\alpha = 1, z_{x, y} = 1, c = 1]$ & $\pmb{-5.94}$ & $\pmb{-6.75}$ & $\pmb{-5.26}$ & $\pmb{-8.23}$ & $-6.45$ & $-8.53$ & $-7.11$ & $-7.94$ & $-6.75$ & $-4.70$ & $-3.34$ & $-2.22$ & $\pmb{-6.90}$ & $-5.93$\\
\hspace{1em}IComb$[\alpha = 1, z_{x} = 1, c = 1]$ & $-2.51$ & $-0.36$ & $-3.97$ & $-7.09$ & $\pmb{-8.22}$ & $\pmb{-9.85}$ & $\pmb{-10.15}$ & $\pmb{-8.81}$ & $\pmb{-11.52}$ & $-6.96$ & $\pmb{-8.43}$ & $\pmb{-8.96}$ & $-5.52$ & $\pmb{-7.52}$\\
\hspace{1em}IComb$[\alpha = 1, z_{x, y} = 0, c = 1]$ & $-3.92$ & $-2.69$ & $-3.88$ & $-6.69$ & $-7.89$ & $-8.69$ & $-8.16$ & $-7.06$ & $-8.97$ & $-7.38$ & $-7.58$ & $-7.82$ & $-5.75$ & $-6.91$\\
\hspace{1em}IComb$[\alpha = 0, z_{x, y} = 1, c = 1]$ & $-1.16$ & $-3.50$ & $-2.93$ & $-5.59$ & $-4.29$ & $-4.73$ & $-4.22$ & $-2.57$ & $-2.84$ & $-1.26$ & $-0.19$ & $-0.41$ & $-3.77$ & $-2.67$\\
\hspace{1em}IComb$[\alpha = 0, z_{x} = 1, c = 1]$ & $-1.35$ & $-3.39$ & $-2.86$ & $-5.72$ & $-4.36$ & $-4.72$ & $-4.21$ & $-2.33$ & $-2.80$ & $-1.27$ & $-0.22$ & $-0.23$ & $-3.80$ & $-2.64$\\
\hspace{1em}IComb$[\alpha = 0, z_{x, y} = 0, c = 1]$ & $-4.13$ & $-3.17$ & $-4.21$ & $-6.08$ & $-7.64$ & $-7.73$ & $-7.89$ & $-7.68$ & $-9.39$ & $\pmb{-7.89}$ & $-8.15$ & $-8.11$ & $-5.59$ & $-7.04$\\
\hspace{1em}Base ($\times 10^4$) & {\it 348.24} & {\it 351.99} & {\it 372.29} & {\it 395.52} & {\it 395.79} & {\it 404.53} & {\it 428.94} & {\it 438.64} & {\it 464.60} & {\it 501.05} & {\it 514.11} & {\it 542.09} & {\it 378.06} & {\it 429.82}\\
\cmidrule{2-15}
\multicolumn{15}{l}{\it States}\\
\hspace{1em}BU & $10.34$ & $7.36$ & $8.70$ & $13.20$ & $12.53$ & $11.08$ & $11.41$ & $14.68$ & $13.83$ & $20.13$ & $17.61$ & $19.31$ & $10.55$ & $13.59$\\
\hspace{1em}OLS & $-2.55$ & $-1.78$ & $-0.59$ & $1.22$ & $0.93$ & $-0.24$ & $-0.58$ & $-0.53$ & $-0.19$ & $0.89$ & $0.28$ & $0.67$ & $-0.49$ & $-0.17$\\
\hspace{1em}WLS$_v$ & $1.91$ & $1.03$ & $2.01$ & $5.04$ & $4.49$ & $3.23$ & $4.03$ & $5.67$ & $5.01$ & $8.79$ & $7.34$ & $8.30$ & $2.96$ & $4.89$\\
\hspace{1em}MinT & $-0.26$ & $-0.88$ & $-0.04$ & $2.59$ & $2.18$ & $0.81$ & $1.45$ & $2.55$ & $1.68$ & $5.77$ & $4.55$ & $5.33$ & $0.74$ & $2.27$\\
\hspace{1em}EMinTU & $925.42$ & $1324.17$ & $1683.98$ & $1990.22$ & $2748.09$ & $2804.73$ & $3128.05$ & $3409.19$ & $3535.36$ & $3692.96$ & $3895.42$ & $3789.60$ & $1924.69$ & $2809.33$\\
\hspace{1em}IComb$[\alpha = 1, z_{x, y} = 1, c = 1]$ & $-7.69$ & $\pmb{-6.91}$ & $-4.30$ & $-1.80$ & $-2.68$ & $-5.49$ & $-5.28$ & $-4.03$ & $-4.98$ & $-2.64$ & $-3.37$ & $-1.46$ & $-4.80$ & $-4.15$\\
\hspace{1em}IComb$[\alpha = 1, z_{x} = 1, c = 1]$ & $-5.06$ & $-3.10$ & $-4.11$ & $-1.37$ & $-4.60$ & $\pmb{-7.54}$ & $\pmb{-8.10}$ & $-5.75$ & $-7.13$ & $-3.83$ & $-4.97$ & $-5.75$ & $-4.32$ & $-5.16$\\
\hspace{1em}IComb$[\alpha = 1, z_{x, y} = 0, c = 1]$ & $-8.03$ & $-5.49$ & $-4.69$ & $-3.11$ & $-4.49$ & $-6.05$ & $-6.54$ & $-6.11$ & $-7.69$ & $-5.46$ & $-6.34$ & $-5.93$ & $-5.31$ & $-5.86$\\
\hspace{1em}IComb$[\alpha = 0, z_{x, y} = 1, c = 1]$ & $-6.74$ & $-6.86$ & $-5.57$ & $-2.34$ & $-2.99$ & $-5.29$ & $-5.70$ & $-3.28$ & $-4.54$ & $-0.78$ & $-1.83$ & $-1.58$ & $-4.96$ & $-3.86$\\
\hspace{1em}IComb$[\alpha = 0, z_{x} = 1, c = 1]$ & $-6.87$ & $-6.79$ & $-5.53$ & $-2.34$ & $-3.02$ & $-5.35$ & $-5.70$ & $-3.12$ & $-4.49$ & $-0.79$ & $-1.85$ & $-1.50$ & $-4.98$ & $-3.85$\\
\hspace{1em}IComb$[\alpha = 0, z_{x, y} = 0, c = 1]$ & $\pmb{-8.67}$ & $-6.63$ & $\pmb{-6.00}$ & $\pmb{-3.60}$ & $\pmb{-5.08}$ & $-6.65$ & $-7.68$ & $\pmb{-7.04}$ & $\pmb{-8.49}$ & $\pmb{-5.61}$ & $\pmb{-6.88}$ & $\pmb{-6.67}$ & $\pmb{-6.10}$ & $\pmb{-6.61}$\\
\hspace{1em}Base ($\times 10^4$) & {\it 25.88} & {\it 25.92} & {\it 26.29} & {\it 26.19} & {\it 26.44} & {\it 27.57} & {\it 28.75} & {\it 28.67} & {\it 29.77} & {\it 29.88} & {\it 31.32} & {\it 32.26} & {\it 26.38} & {\it 28.25}\\
\cmidrule{2-15}
\multicolumn{15}{l}{\it Zones}\\
\hspace{1em}BU & $0.94$ & $1.13$ & $1.46$ & $2.41$ & $2.51$ & $1.96$ & $1.11$ & $2.31$ & $2.48$ & $5.73$ & $3.57$ & $3.70$ & $1.74$ & $2.49$\\
\hspace{1em}OLS & $-3.81$ & $-2.43$ & $-2.46$ & $-2.69$ & $-2.82$ & $-2.84$ & $-3.57$ & $-4.14$ & $-3.75$ & $-3.97$ & $-4.28$ & $-4.56$ & $-2.84$ & $-3.47$\\
\hspace{1em}WLS$_v$ & $-2.78$ & $-1.98$ & $-1.86$ & $-1.79$ & $-2.00$ & $-2.25$ & $-2.67$ & $-2.55$ & $-2.56$ & $-1.32$ & $-2.29$ & $-2.37$ & $-2.11$ & $-2.20$\\
\hspace{1em}MinT & $-3.41$ & $-2.64$ & $-2.66$ & $-2.71$ & $-2.88$ & $-3.06$ & $-3.67$ & $-3.78$ & $-3.80$ & $-2.50$ & $-3.42$ & $-3.52$ & $-2.90$ & $-3.18$\\
\hspace{1em}EMinTU & $1144.09$ & $1561.37$ & $2094.00$ & $2396.67$ & $2924.76$ & $3327.26$ & $3653.74$ & $4148.44$ & $4283.72$ & $4669.22$ & $4855.42$ & $4647.70$ & $2251.60$ & $3363.72$\\
\hspace{1em}IComb$[\alpha = 1, z_{x, y} = 1, c = 1]$ & $\pmb{-10.08}$ & $-8.08$ & $-7.54$ & $-6.35$ & $-6.63$ & $-8.68$ & $-9.81$ & $\pmb{-9.94}$ & $-9.09$ & $\pmb{-8.40}$ & $-8.84$ & $-8.57$ & $-7.89$ & $-8.52$\\
\hspace{1em}IComb$[\alpha = 1, z_{x} = 1, c = 1]$ & $-5.86$ & $-3.56$ & $-4.21$ & $-3.39$ & $-4.74$ & $-6.58$ & $-7.61$ & $-8.23$ & $-7.38$ & $-6.20$ & $-6.74$ & $-7.50$ & $-4.74$ & $-6.06$\\
\hspace{1em}IComb$[\alpha = 1, z_{x, y} = 0, c = 1]$ & $-3.20$ & $-0.53$ & $-0.90$ & $-0.43$ & $-1.43$ & $-2.63$ & $-3.63$ & $-4.72$ & $-4.21$ & $-3.76$ & $-4.52$ & $-4.53$ & $-1.53$ & $-2.94$\\
\hspace{1em}IComb$[\alpha = 0, z_{x, y} = 1, c = 1]$ & $-9.65$ & $\pmb{-8.21}$ & $\pmb{-8.25}$ & $-7.41$ & $-7.74$ & $-9.15$ & $-10.24$ & $-9.67$ & $\pmb{-9.23}$ & $-7.92$ & $-8.84$ & $\pmb{-9.03}$ & $-8.41$ & $\pmb{-8.79}$\\
\hspace{1em}IComb$[\alpha = 0, z_{x} = 1, c = 1]$ & $-9.74$ & $-8.18$ & $\pmb{-8.25}$ & $\pmb{-7.42}$ & $\pmb{-7.78}$ & $\pmb{-9.17}$ & $\pmb{-10.27}$ & $-9.61$ & $-9.20$ & $-7.88$ & $-8.85$ & $-8.97$ & $\pmb{-8.42}$ & $\pmb{-8.79}$\\
\hspace{1em}IComb$[\alpha = 0, z_{x, y} = 0, c = 1]$ & $-7.73$ & $-5.48$ & $-5.69$ & $-5.42$ & $-6.17$ & $-7.09$ & $-8.21$ & $-9.04$ & $-8.65$ & $-7.94$ & $\pmb{-8.98}$ & $-8.99$ & $-6.27$ & $-7.50$\\
\hspace{1em}Base ($\times 10^4$) & {\it 4.52} & {\it 4.46} & {\it 4.51} & {\it 4.56} & {\it 4.59} & {\it 4.69} & {\it 4.82} & {\it 4.85} & {\it 4.88} & {\it 4.94} & {\it 5.09} & {\it 5.19} & {\it 4.55} & {\it 4.76}\\
\cmidrule{2-15}
\multicolumn{15}{l}{\it Regions}\\
\hspace{1em}BU & $0.00$ & $0.00$ & $0.00$ & $0.00$ & $0.00$ & $0.00$ & $0.00$ & $0.00$ & $0.00$ & $0.00$ & $0.00$ & $0.00$ & $0.00$ & $0.00$\\
\hspace{1em}OLS & $-2.17$ & $-1.96$ & $-2.06$ & $-2.69$ & $-2.61$ & $-2.42$ & $-2.38$ & $-3.20$ & $-2.99$ & $-4.50$ & $-3.71$ & $-3.77$ & $-2.32$ & $-2.90$\\
\hspace{1em}WLS$_v$ & $-1.92$ & $-1.73$ & $-1.84$ & $-2.21$ & $-2.39$ & $-2.26$ & $-2.13$ & $-2.84$ & $-2.80$ & $-3.94$ & $-3.50$ & $-3.23$ & $-2.06$ & $-2.59$\\
\hspace{1em}MinT & $-2.41$ & $-2.17$ & $-2.39$ & $-2.82$ & $-3.02$ & $-2.89$ & $-2.79$ & $-3.59$ & $-3.43$ & $-4.66$ & $-4.19$ & $-3.86$ & $-2.62$ & $-3.21$\\
\hspace{1em}EMinTU & $1224.69$ & $1623.45$ & $2007.34$ & $2460.61$ & $3041.47$ & $3539.59$ & $3838.63$ & $4348.14$ & $4545.91$ & $4893.62$ & $5186.61$ & $4888.86$ & $2325.00$ & $3516.26$\\
\hspace{1em}IComb$[\alpha = 1, z_{x, y} = 1, c = 1]$ & $\pmb{-9.20}$ & $-7.59$ & $-7.82$ & $-7.71$ & $-7.52$ & $-8.36$ & $-8.21$ & $-9.61$ & $-8.54$ & $-9.84$ & $-8.35$ & $-8.66$ & $-8.04$ & $-8.47$\\
\hspace{1em}IComb$[\alpha = 1, z_{x} = 1, c = 1]$ & $-5.49$ & $-3.44$ & $-3.99$ & $-4.32$ & $-4.68$ & $-5.27$ & $-5.50$ & $-6.81$ & $-5.94$ & $-6.98$ & $-5.31$ & $-5.48$ & $-4.54$ & $-5.30$\\
\hspace{1em}IComb$[\alpha = 1, z_{x, y} = 0, c = 1]$ & $-1.38$ & $0.19$ & $-0.16$ & $-0.03$ & $-0.56$ & $-1.11$ & $-1.17$ & $-2.76$ & $-1.80$ & $-3.15$ & $-1.30$ & $-1.10$ & $-0.52$ & $-1.22$\\
\hspace{1em}IComb$[\alpha = 0, z_{x, y} = 1, c = 1]$ & $-9.17$ & $\pmb{-7.86}$ & $\pmb{-8.25}$ & $\pmb{-8.26}$ & $-8.43$ & $\pmb{-8.99}$ & $-9.15$ & $\pmb{-9.71}$ & $\pmb{-8.97}$ & $\pmb{-10.00}$ & $-8.73$ & $\pmb{-9.09}$ & $\pmb{-8.50}$ & $\pmb{-8.90}$\\
\hspace{1em}IComb$[\alpha = 0, z_{x} = 1, c = 1]$ & $\pmb{-9.20}$ & $-7.83$ & $-8.24$ & $-8.25$ & $\pmb{-8.47}$ & $\pmb{-8.99}$ & $\pmb{-9.16}$ & $-9.67$ & $\pmb{-8.97}$ & $-9.97$ & $\pmb{-8.74}$ & $-9.05$ & $\pmb{-8.50}$ & $-8.89$\\
\hspace{1em}IComb$[\alpha = 0, z_{x, y} = 0, c = 1]$ & $-7.04$ & $-5.47$ & $-5.85$ & $-6.15$ & $-6.60$ & $-6.92$ & $-6.88$ & $-8.13$ & $-7.48$ & $-8.74$ & $-7.53$ & $-7.55$ & $-6.35$ & $-7.06$\\
\hspace{1em}Base ($\times 10^4$) & {\it 1.36} & {\it 1.32} & {\it 1.34} & {\it 1.35} & {\it 1.37} & {\it 1.39} & {\it 1.41} & {\it 1.43} & {\it 1.43} & {\it 1.48} & {\it 1.48} & {\it 1.50} & {\it 1.35} & {\it 1.40}\\
\cmidrule{2-15}
\multicolumn{15}{l}{\it Average}\\
\hspace{1em}BU & $19.86$ & $15.37$ & $15.41$ & $16.41$ & $17.58$ & $18.68$ & $19.04$ & $23.38$ & $22.94$ & $24.93$ & $24.32$ & $25.41$ & $17.23$ & $20.60$\\
\hspace{1em}OLS & $-1.49$ & $-1.23$ & $-1.24$ & $-1.12$ & $-1.19$ & $-1.17$ & $-1.12$ & $-1.08$ & $-1.08$ & $-1.01$ & $-1.01$ & $-0.98$ & $-1.24$ & $-1.13$\\
\hspace{1em}WLS$_v$ & $7.87$ & $5.82$ & $5.34$ & $5.29$ & $5.75$ & $6.87$ & $8.05$ & $10.20$ & $9.70$ & $10.69$ & $10.55$ & $11.08$ & $6.15$ & $8.29$\\
\hspace{1em}MinT & $5.00$ & $3.17$ & $2.55$ & $2.10$ & $2.72$ & $3.56$ & $4.65$ & $6.20$ & $5.55$ & $6.74$ & $6.85$ & $6.90$ & $3.17$ & $4.81$\\
\hspace{1em}EMinTU & $940.75$ & $1332.73$ & $1773.83$ & $2012.41$ & $2543.20$ & $2771.72$ & $3187.61$ & $3400.82$ & $3404.38$ & $3643.60$ & $3670.06$ & $3652.28$ & $1917.64$ & $2778.11$\\
\hspace{1em}IComb$[\alpha = 1, z_{x, y} = 1, c = 1]$ & $\pmb{-7.48}$ & $\pmb{-7.12}$ & $\pmb{-5.72}$ & $\pmb{-6.41}$ & $-5.75$ & $-7.83$ & $-7.23$ & $-7.56$ & $-6.90$ & $-5.38$ & $-4.69$ & $-3.65$ & $\pmb{-6.72}$ & $-6.23$\\
\hspace{1em}IComb$[\alpha = 1, z_{x} = 1, c = 1]$ & $-4.08$ & $-1.94$ & $-4.04$ & $-4.87$ & $-6.40$ & $\pmb{-8.23}$ & $\pmb{-8.72}$ & $-7.78$ & $\pmb{-9.25}$ & $-6.17$ & $-7.06$ & $-7.66$ & $-4.99$ & $-6.48$\\
\hspace{1em}IComb$[\alpha = 1, z_{x, y} = 0, c = 1]$ & $-4.44$ & $-2.63$ & $-3.12$ & $-4.06$ & $-5.17$ & $-6.18$ & $-6.23$ & $-5.96$ & $-7.13$ & $-5.95$ & $-6.15$ & $-6.19$ & $-4.30$ & $-5.37$\\
\hspace{1em}IComb$[\alpha = 0, z_{x, y} = 1, c = 1]$ & $-4.98$ & $-5.65$ & $-5.08$ & $-5.48$ & $-5.06$ & $-6.08$ & $-6.08$ & $-4.68$ & $-4.88$ & $-3.12$ & $-2.75$ & $-2.83$ & $-5.39$ & $-4.63$\\
\hspace{1em}IComb$[\alpha = 0, z_{x} = 1, c = 1]$ & $-5.11$ & $-5.57$ & $-5.04$ & $-5.54$ & $-5.11$ & $-6.09$ & $-6.08$ & $-4.51$ & $-4.85$ & $-3.12$ & $-2.77$ & $-2.70$ & $-5.42$ & $-4.61$\\
\hspace{1em}IComb$[\alpha = 0, z_{x, y} = 0, c = 1]$ & $-6.20$ & $-4.68$ & $-5.08$ & $-5.42$ & $\pmb{-6.70}$ & $-7.27$ & $-7.76$ & $\pmb{-7.79}$ & $-8.85$ & $\pmb{-7.50}$ & $\pmb{-7.91}$ & $\pmb{-7.85}$ & $-5.91$ & $\pmb{-7.01}$\\
\hspace{1em}Base ($\times 10^4$) & {\it 6.75} & {\it 6.75} & {\it 6.97} & {\it 7.20} & {\it 7.23} & {\it 7.42} & {\it 7.76} & {\it 7.86} & {\it 8.17} & {\it 8.55} & {\it 8.79} & {\it 9.14} & {\it 7.05} & {\it 7.71}\\
\bottomrule
\end{tabular}}
\end{table}

\section{Conclusion}
In this paper, we show mathematically that if a set of target variables satisfies a set of linear constraints, then their predicted values generated from a multivariate regression model or a multivariate penalised regression model also satisfy the same set of constraints. We demonstrate this in the context of hierarchical forecasting, where the target variables satisfy a set of aggregation constraints. 

The importance of this result in the context of hierarchical forecasting is that when a historical dataset of actual values and their base forecasts, produced by possibly different forecasters, is available but may not satisfy the aggregation constraints, we can frame the problem as finding the best way to extract relevant signals from these individual forecasts to improve the forecasts of all variables in the hierarchy using multiple penalised regression methods. If we do so, our mathematical result guarantees that these forecasts satisfy the aggregation constraints. This forms the basis of the information combination (IComb) methods proposed in this paper.

We follow the existing literature in hierarchical forecasting and use the Australian tourism dataset, along with a set of base forecasts produced by univariate time series forecasting methods, to evaluate our IComb method. We find that IComb produces superior forecasts on many occasions relative to existing forecast reconciliation methods. Among the IComb variants with different penalties, we find that the best-performing models tend to be quite dense in this application. In fact, in many cases, ridge regression, which uses all base forecasts, produces the most accurate reconciled forecasts. We interpret this as evidence that, in this context, each univariate time series forecast contains small signals about all variables in the hierarchy, and therefore dense models are more suitable for combining these weak signals than parsimonious models.

The main message of our paper is that the information combination perspective provides a comprehensive and logical framework for forecast reconciliation in hierarchical time series. It does not rely on assumptions such as the unbiasedness of base forecasts, which are unlikely to hold in practice, and it can readily accommodate situations where multiple sets of forecasts are available at some or all levels of the hierarchy. Moreover, the application to the Australian tourism dataset affirms its potential to deliver substantial improvements in forecast accuracy, more so than several other reconciliation methods. 

In the process of this research, we identified a couple of issues that are worth mentioning. First, we realised that the statement ``top-down reconciliation can never produce unbiased forecasts even if the base forecasts are unbiased'', which is frequently repeated in the hierarchical forecasting literature, is incorrect. We provide a simple counterexample in the main body of the paper and include an appendix explaining the basis of this misconception. Second, we found that the default method for determining the minimum value of the penalty factor in \texttt{cv.glmnet} produced a relatively large minimum in our empirical application, resulting in a very narrow search over possible penalty factors. We therefore developed an alternative rule for determining the minimum penalty factor in our cross-validation exercises.

\section{Acknowledgments}

The authors wish to acknowledge the use of Monash eResearch capabilities, including MonARCH, for high-performance computing.

\textit{Conflicts of interest}: The authors declare no conflict of interest.

\section{Data availability}

Dataset used in this paper is publicly available. All coding was done in R. Our programs are available in the \href{https://github.com/ShanikaLW/icomb-paper}{public domain} for readers interested in replicating our results or applying our methods to their own data.

\clearpage

\printbibliography

\clearpage

\section*{Appendices}
\appendix

\section[On the sufficiency of S G S = S]{On the sufficiency of $\Sb \Gb \Sb = \Sb$}\label{app:A}

\citet{WicEtAl2019} state that if the base forecasts $\yhat_{T+h|T}$ are conditionally unbiased, as defined in equation (\ref{eq:unbiased}), the reconciled forecasts $\ytilde_{T+h|h} = \Sb \Gb \yhat_{T+h|T}$ will be conditionally unbiased if and only if $\Sb \Gb \Sb = \Sb$. The reasoning is based on the following steps:
\begin{align*}
	\yb_{T+h} = \Sb \bb_{T+h} & \Rightarrow E(\yb_{T+h} \mid \mathcal{I}_T) = \Sb E(\bb_{T+h} \mid \mathcal{I}_T) \\
	\ytilde_{T+h|T} \text{ unbiased} & \Rightarrow E(\ytilde_{T+h|T} \mid \mathcal{I}_T) = E(\yb_{T+h} \mid \mathcal{I}_T) = \Sb E(\bb_{T+h} \mid \mathcal{I}_T) \\
	\yhat_{T+h|T} \text{ unbiased} & \Rightarrow E(\ytilde_{T+h|T} \mid \mathcal{I}_T) = \Sb \Gb E(\yhat_{T+h|T} \mid \mathcal{I}_T) = \Sb \Gb E(\yb_{T+h} \mid \mathcal{I}_T) = \Sb \Gb \Sb E(\bb_{T+h} \mid \mathcal{I}_T) 
\end{align*} 
The last two lines show that conditional unbiasedness implies 
\begin{equation}
\Sb E(\bb_{T+h} \mid \mathcal{I}_T) = \Sb \Gb \Sb E(\bb_{T+h} \mid \mathcal{I}_T). \label{eq:appA}
\end{equation}

The condition $\Sb \Gb \Sb = \Sb$ is sufficient for this equality to hold, but it is not necessary. If $E(\bb_{T+h} \mid \mathcal{I}_T)$ belongs to a specific subspace of $\mathbb{R}^n$ with dimension $r < n$, then we can write $E(\bb_{T+h} \mid \mathcal{I}_T) = \bm{A} \bm{\beta}(\mathcal{I}_T)$, where $\bm{A}$ is an $n \times r$ matrix and $\bm{\beta}(\mathcal{I}_T)$ is an $r \times 1$ vector function of the history\footnote{Refer to Section \ref{sec:sim} for an example.}. In this case, from equation (\ref{eq:appA}) we see that $\Sb \Gb \Sb \bm{A}= \Sb \bm{A}$ would be necessary. While $\Sb \Gb \Sb = \Sb$ implies $\Sb \Gb \Sb \bm{A}= \Sb \bm{A}$, the latter does not necessarily imply $\bm{SGS} = \bm{S}$. In the context of forecasting disaggregated time series, it is quite conceivable that at least some components may be generated by a dynamic factor model, in which case their conditional expectations will be linear combinations of conditional expectations of a small number of common dynamic factors.  

\section{Proof of Proposition \ref{prop:ml}}\label{app:mlasso}

The multivariate lasso estimator $\hat{\bm{B}}^{*\text{ml}}$ minimises the following criterion:
\begin{align*}
\frac{1}{2T} \|\bm{Y}^* -\bm{X}^*\bm{B}^* \|_2^2 + \lambda\sum_{j=1}^k\|\bm{\beta}^*_j\|_2 = \frac{1}{2T} \|\bm{Y}^* -\bm{X}^*_{:, -i}\bm{B}^*_{-i, :} - \bm{x}^*_i\bm{\beta}^{*'}_i\|_2^2 + \lambda\sum_{j=1}^k\|\bm{\beta}^*_j\|_2,
\end{align*}
where $\bm{\beta}^*_j$ is the transpose of the $j$-th row of $\bm{B}^*$, i.e., the $j$-th column of $\bm{B}^{*'}$, $\bm{x}^*_i$ is the $i$-th column of $\bm{X}^*$, and $\bm{X}^*_{:,-i}$ and $\bm{B}^*_{-i,:}$ denote two matrices obtained by removing the $i$-th column of $\bm{X}^*$ and $i$-th row of $\bm{B}^*$, respectively.

The Karush-Kuhn-Tucker condition for $\hat{\bm{B}}^{*\text{ml}}$ to be a solution to the above minimisation problem is \citep{YuaLim2006} 
\begin{align*}
-\frac{1}{T}[\bm{Y}^* -\bm{X}^*_{:, -i}\hat{\bm{B}}^*_{-i, :} - \bm{x}^*_i\hat{\bm{\beta}}^{*'}_i]'\bm{x}^*_i + \lambda\frac{\hat{\bm{\beta}}^*_i}{\|\hat{\bm{\beta}}^*_i\|_2}	= \bm{0}, \qquad \text{for}\ \hat{\bm{\beta}}^*_i \neq \bm{0},
\end{align*}
for $i = 1, 2, \dots, r$, where $r (\leq k)$ is the number of non-zero groups (active predictors).

Stacking all $r$ subgradient conditions next to each other, we obtain

\begin{equation}
-\frac{1}{T}(\bm{Y}^* - \bm{X}^*_a\hat{\bm{B}}^*_a)'\bm{X}^*_a + \hat{\bm{B}}^{*'}_a\bm{\Lambda} = \bm{0}, \label{eq:KKT}
\end{equation}

where $\bm{X}^*_a$ contains the active $r$ columns of $\bm{X}^*$ for which the corresponding $\hat{\bm{\beta}}^*_i\neq \bm{0}\ (i \in \{1, 2, \dots, k\})$, $\bm{\Lambda}$ is an $r \times r$ diagonal matrix whose $i$-th element is $\frac{\lambda}{\|\hat{\bm{\beta}}^*_i\|_2}$ and $\hat{\bm{B}}^*_a$ is the estimated non-zero coefficient matrix. 

Let $\hat{\bm{B}}_a = \hat{\bm{\Sigma}}^{-1/2}_{x, a} \hat{\bm{B}}^*_a\hat{\bm{\Sigma}}^{1/2}_y$, where $\hat{\bm{\Sigma}}_{x,a}$ is a submatrix of $\hat{\bm{\Sigma}}_x$ consisting of the variances of $\bm{X}_a^*$. Equation (\ref{eq:KKT}) implies that
\begin{align*}
& \hat{\bm{B}}^{*'}_a = \bm{Y}^{*'}\bm{X}^*_a(\bm{X}^{*'}_a\bm{X}^*_a + T\bm{\Lambda})^{-1}\\
\Rightarrow & \bm{S}'_{\bot}\hat{\bm{B}}'_a = \bm{0} \ \text{because}\ \bm{S}'_{\bot}\bm{Y} = \bm{0},
\end{align*}
which shows that the multivariate lasso estimates of the coefficients of the active predictors in different equations satisfy the aggregation constraints. Since $\bm{\hat{B}}^{\text{ml}'}$ is the same as $\bm{\hat{B}}'_a$ with some zero columns inserted into it for inactive predictors, it follows that 
\begin{equation*}
    \Sb_\perp ' \bm{\hat{B}}^{\text{ml}'} = \bm{0},
\end{equation*}
which completes the proof.

\section{A new interpretation of the MinT sample reconciliation method}\label{app:MinT}

\citet{WicEtAl2019} translate the question of forecast reconciliation to the problem of finding the $n \times m$ matrix $\Gb$ in $\ytilde_{T+h|T} = \Sb \Gb \yhat_{T+h|T}$ such that the trace of the variance-covariance matrix of the reconciled forecast errors is minimised subject to $\Sb \Gb \Sb = \Sb$. They derive the optimal $\Gb$,
\begin{equation}
\Gb = (\Sb' \Wb_h^{-1} \Sb)^{-1} \Sb'{\Wb_h}^{-1}, \label{eq:opremat1}
\end{equation}
where $\Wb$ is the positive-definite covariance matrix of the $h$-step ahead base forecast errors and show that it can be equivalently computed using
\begin{equation}
	\Gb = \Jb - \Jb \Wb_h \Sb_\perp ( \Sb_\perp'\Wb_h \Sb_\perp)^{-1} \Sb_\perp', \label{eq:opremat2}
\end{equation}
where
\begin{equation*}
	\begin{matrix}
		\underset{n \times m}{\Sb'} = \left[
			\Cb'  \;\;\;\;  \bm{I}_n \right] \text{,}\;\; & 
			\underset{n \times m}{\Jb} = \left[ 
			\underset{n \times (m-n)}{\bm{0}}  \;\;\;\; \bm{I}_n
		 \right] \text{,}\;\; &
			\underset{(m-n) \times m}{\Sb_\perp'} = 
		\left[ 	\bm{I}_{m-n}  \;\;\;\; -\Cb
		\right].
	\end{matrix}
\end{equation*}

They use the notation $\bm{U}$ for $\Sb_\perp$, but we prefer $\Sb_\perp$ to make it clear that $\Sb_\perp'\Sb = \bm{0}$.
Given that the total number of series in the hierarchy, $m$, and the number of series at the bottom level, $n$, are likely to be significantly greater than the number of non-bottom level series in the hierarchy, $m - n$, \citet{WicEtAl2019} state that using the formulation in equation (\ref{eq:opremat2}), which only requires the inversion of the $(m-n) \times (m-n)$ matrix $\Sb_\perp'\Wb_h \Sb_\perp$, is more computationally efficient than using the formulation in equation (\ref{eq:opremat1}), which requires the inversion of two larger matrices, namely the ${m \times m}$ matrix $\Wb_h$ and the $n \times n$ matrix $\Sb' \Wb_h^{-1} \Sb$. 

Without loss of generality, we consider the case $h=1$, and we drop the conditional set from the notation of forecasts and forecast errors for simplicity. That is, in what follows, $\yhat_{t}$ denotes the one-step-ahead base forecast of $\yb_{t}$ and $\hat{\eb}_{t} = \yb_{t} - \yhat_{t}$. The MinT sample method uses the information on observed values and forecasts available up to time $T$ to form
\[
\hat{\Wb}_1 = \frac{1}{T}\sum_{t=1}^{T}{\hat{\bm{e}}_t\hat{\bm{e}}'_t}= \frac{1}{T} \Eb'\Eb, 
\]
where $\Eb' = \begin{bmatrix}
	{\hat{\eb}_1} & {\hat{\eb}_2} & \hdots &
	{\hat{\eb}_{T}}
\end{bmatrix}$.
Substituting this in the formulation of $\Gb$ in equation (\ref{eq:opremat2}), we obtain
\begin{equation}
	\Gb = \Jb - \Jb \Eb'\Eb \Sb_\perp ( \Sb_\perp'\Eb'\Eb \Sb_\perp)^{-1} \Sb_\perp'. \label{eq:opremat3}
\end{equation} 

We note that
\begin{align*}
	\Jb \Eb' & = [ \underset{n \times (m-n)}{\bm{0}}  \;\;\;\; \bm{I}_n ] \left[
		{\hat{\eb}_1} \;\; {\hat{\eb}_2} \;\; \hdots \;\;
		{\hat{\eb}_{T}}\right] \\
		& = \left[{\hat{\eb}^{bot}_1} \;\; {\hat{\eb}^{bot}_2} \;\; \hdots \;\;
		{\hat{\eb}^{bot}_{T}}\right] := (\Eb^{bot})',
\end{align*}
where $\hat{\eb}^{bot}_{t}$ is the error of the one-step-ahead base forecasts of bottom-level series at time $t$.

Let $\hat{\Yb}' = \begin{bmatrix}
	\hat{\yb}_1 & \hat{\yb}_2 & \hdots &
	\hat{\yb}_{T}
\end{bmatrix}$.
We can derive
\begin{align*}
	\Sb_\perp' \Eb' & = \Sb_\perp' (\Yb' - \hat{\Yb}') \\
	& = -\Sb_\perp' \hat{\Yb}' \;\;\;\;(\text{because } \Sb_\perp' \Yb' =\bm{0}) \\
	&= -\left[\bm{I}_{m-n}  \;\;\;\; -\Cb \right] \left[\yhat_1\;\;\yhat_2\;\;\ldots \;\;\yhat_{T}\right] \\
	& =	 -\left[ (\yhat_1^{top}-\Cb \yhat_1^{bot}) \;\;\;(\yhat_2^{top}-\Cb \yhat_2^{bot})\;\;\ldots \;\;(\yhat^{top}_{T} - \Cb \yhat^{bot}_{T})\right] := -\Zb',
\end{align*}
where $\yhat^{bot}_{t}$ is the one-step-ahead base forecasts of the $n$ bottom-level series at time $t$, and $\yhat^{top}_{t}$ are the forecasts of the remaining $m-n$ series in the hierarchy at time $t$. Note that $\yhat_t^{top}-\Cb \yhat_t^{bot}$ is the incoherency in forecasts at time $t$, i.e. how far the top-level forecasts were from the corresponding aggregates of the bottom-level forecasts, and the matrix $\Zb$ is the $T \times (m-n)$ matrix of historical incoherencies. Using this notation, we obtain
\begin{align*}
	\Gb = \Jb + (\Eb^{bot})' \Zb ( \Zb'\Zb)^{-1} \Sb_\perp' = \Jb + \bm{\Psi}' \Sb_\perp',
\end{align*} 
where $\bm{\Psi}$ is the least square estimate of the parameters of the multivariate regression of the historical forecast errors of bottom-level forecasts on the incoherency in bottom-level base forecasts. Using this mapping matrix, it becomes clear that the coherent forecasts of the bottom-level series in the MinT sample procedure are
\begin{equation*}
	\ytilde_{T+h}^{bot} = \Gb \yhat_{T+h} = \Jb\yhat_{T+h} + \bm{\Psi}' \Sb_\perp'\yhat_{T+h} = \yhat_{T+h}^{bot} + \bm{\Psi}' (\yhat^{top}_{T+h} - \Cb \yhat^{bot}_{T+h})
\end{equation*}
leading to
\begin{equation*}
	\ytilde_{T+h} = \Sb \Gb \yhat_{T+h} = \Sb\ytilde_{T+h}^{bot}.
\end{equation*}
This makes it clear that the MinT procedure only allows the forecast to be modified on the basis of incoherencies. It determines, from historical data, how much can be learnt from these coherencies, and modifies the new forecasts accordingly. This makes sense given that MinT's conditional unbiasedness assumption implies that there is nothing in the historical information set that can help improve forecasts, and the only thing that the procedure needs to do is to fix their incoherency. Hence the procedure finds out what part of the historical forecast errors are due to these incoherencies, and adjusts future forecasts accordingly.

This angle of looking at the MinT procedure suggests an alternative method regarding shrinkage. So far in the literature, the shrinkage procedures suggested for MinT are all based on shrinking the elements of the $\hat{\Wb}_1$ matrix, which is a symmetric $m \times m$ matrix. The above view of the MinT procedure reveals that we only need to consider shrinkage of the parameter estimates in regressions with $m-n$ regressors, i.e. shrinking the $ \Zb'\Zb$ matrix that is a symmetric $(m-n) \times (m-n)$ matrix. This is akin to the important insight of \citet{White1980} that for proper inference about the ordinary least squares estimator in a multiple linear regression with $k$ parameters and non-spherical errors, one need not estimate the variance-covariance matrix of the errors $\bm{\Omega}$, rather the only thing needed is an estimate of the $k\times k$ matrix $\Xb' \bm{\Omega}\Xb$. 

\section{Forecasting performance of tourism data set}
\label{app:all-results}

\subsection{ARIMA forecasts}

\begin{footnotesize}
\begin{landscape}
\begin{longtable}[h]{lrrrrrrrrrrrrrr}
\caption{\mbox{PRIAL for forecast reconciliation methods using ARIMA base forecasts.}}\\
\toprule
& \multicolumn{14}{c}{Forecast horizon $(h)$}\\
\cmidrule{2-15}
  & {\it 1} & {\it 2} & {\it 3} & {\it 4} & {\it 5} & {\it 6} & {\it 7} & {\it 8} & {\it 9} & {\it 10} & {\it 11} & {\it 12} & {\it 1--6} & {\it 1--12}\\
 \cmidrule{2-15}
\endfirsthead
\caption{(continued from previous page)}\\
\toprule
& \multicolumn{14}{c}{Forecast horizon $(h)$}\\
\cmidrule{2-15}
  & {\it 1} & {\it 2} & {\it 3} & {\it 4} & {\it 5} & {\it 6} & {\it 7} & {\it 8} & {\it 9} & {\it 10} & {\it 11} & {\it 12} & {\it 1--6} & {\it 1--12}\\
 \cmidrule{2-15}
\endhead
\bottomrule
\endfoot
\multicolumn{15}{l}{ \it Australia}\\
\hspace{1em}BU & $53.73$ & $62.64$ & $63.58$ & $52.97$ & $56.61$ & $56.62$ & $62.85$ & $59.57$ & $72.53$ & $60.50$ & $52.84$ & $55.91$ & $57.56$ & $59.06$\\
\hspace{1em}OLS & $0.01$ & $-0.01$ & $0.68$ & $0.02$ & $0.22$ & $-0.05$ & $1.03$ & $0.38$ & $2.37$ & $1.40$ & $0.57$ & $1.16$ & $0.14$ & $0.68$\\
\hspace{1em}WLS$_v$ & $23.46$ & $27.45$ & $31.40$ & $23.51$ & $26.06$ & $25.25$ & $31.01$ & $28.20$ & $38.62$ & $30.26$ & $24.63$ & $27.85$ & $26.12$ & $28.16$\\
\hspace{1em}MinT(Shrink) & $13.86$ & $14.48$ & $20.16$ & $13.27$ & $14.22$ & $12.73$ & $18.68$ & $15.21$ & $24.01$ & $17.82$ & $12.91$ & $16.50$ & $14.71$ & $16.14$\\
\hspace{1em}EMinTU & $362.12$ & $424.60$ & $406.27$ & $324.13$ & $280.36$ & $300.57$ & $304.59$ & $300.86$ & $281.66$ & $285.19$ & $152.46$ & $262.24$ & $346.43$ & $300.35$\\
\hspace{1em}IComb$[\alpha = 1, z_{x, y} = 1, c = 1]$ & $5.73$ & $8.44$ & $15.71$ & $4.67$ & $6.38$ & $5.65$ & $10.02$ & $1.91$ & $13.86$ & $7.31$ & $-1.86$ & $3.18$ & $7.64$ & $6.44$\\
\hspace{1em}IComb$[\alpha = 1, z_{x, y} = 1, c = 0]$ & $4.13$ & $7.31$ & $14.10$ & $4.09$ & $6.68$ & $5.51$ & $11.44$ & $6.61$ & $15.91$ & $9.28$ & $0.82$ & $5.97$ & $6.89$ & $7.52$\\
\hspace{1em}IComb$[\alpha = 1, z_{x} = 1, c = 1]$ & $-2.42$ & $-1.45$ & $12.48$ & $1.40$ & $2.04$ & $-2.85$ & $-1.59$ & $-5.53$ & $5.12$ & $0.97$ & $-5.34$ & $1.29$ & $1.47$ & $0.17$\\
\hspace{1em}IComb$[\alpha = 1, z_{x} = 1, c = 0]$ & $-1.17$ & $0.28$ & $14.13$ & $2.10$ & $1.48$ & $-2.96$ & $0.09$ & $-3.18$ & $5.55$ & $3.08$ & $-4.47$ & $2.93$ & $2.19$ & $1.32$\\
\hspace{1em}IComb$[\alpha = 1, z_{x, y} = 0, c = 1]$ & $-0.90$ & $\pmb{-7.64}$ & $1.72$ & $-4.22$ & $-4.73$ & $-7.79$ & $-7.46$ & $-7.47$ & $-0.33$ & $-3.70$ & $-6.67$ & $-2.14$ & $-4.04$ & $-4.36$\\
\hspace{1em}IComb$[\alpha = 1, z_{x, y} = 0, c = 0]$ & $\pmb{-4.25}$ & $-6.56$ & $-0.16$ & $-2.38$ & $-6.69$ & $-10.29$ & $\pmb{-10.72}$ & $-10.32$ & $-3.16$ & $-7.27$ & $-9.78$ & $-5.29$ & $-5.16$ & $-6.57$\\
\hspace{1em}IComb$[\alpha = 0, z_{x, y} = 1, c = 1]$ & $2.80$ & $5.48$ & $10.09$ & $3.32$ & $3.99$ & $0.94$ & $0.59$ & $-2.65$ & $4.98$ & $-1.80$ & $-9.68$ & $-5.20$ & $4.33$ & $0.56$\\
\hspace{1em}IComb$[\alpha = 0, z_{x, y} = 1, c = 0]$ & $2.28$ & $4.68$ & $10.96$ & $4.28$ & $5.61$ & $3.30$ & $3.30$ & $0.43$ & $9.36$ & $3.40$ & $-5.00$ & $0.42$ & $5.15$ & $3.29$\\
\hspace{1em}IComb$[\alpha = 0, z_{x} = 1, c = 1]$ & $2.96$ & $5.69$ & $10.14$ & $3.35$ & $3.97$ & $0.99$ & $0.48$ & $-2.66$ & $4.87$ & $-1.99$ & $-9.61$ & $-5.15$ & $4.41$ & $0.57$\\
\hspace{1em}IComb$[\alpha = 0, z_{x} = 1, c = 0]$ & $2.28$ & $4.54$ & $10.72$ & $4.26$ & $6.16$ & $2.98$ & $2.74$ & $0.71$ & $9.06$ & $3.39$ & $-4.93$ & $0.24$ & $5.13$ & $3.22$\\
\hspace{1em}IComb$[\alpha = 0, z_{x, y} = 0, c = 1]$ & $-3.17$ & $-5.27$ & $-1.08$ & $-5.46$ & $-6.95$ & $-9.35$ & $-6.96$ & $-10.28$ & $-5.67$ & $-6.66$ & $-8.76$ & $-4.70$ & $-5.36$ & $-6.34$\\
\hspace{1em}IComb$[\alpha = 0, z_{x, y} = 0, c = 0]$ & $-4.07$ & $-6.18$ & $\pmb{-1.26}$ & $\pmb{-5.95}$ & $\pmb{-7.90}$ & $\pmb{-10.61}$ & $-8.73$ & $\pmb{-11.48}$ & $\pmb{-8.20}$ & $\pmb{-9.29}$ & $\pmb{-11.33}$ & $\pmb{-7.27}$ & $\pmb{-6.15}$ & $\pmb{-7.92}$\\
\cmidrule{2-15}
\multicolumn{15}{l}{\it States}\\
\hspace{1em}BU & $21.15$ & $22.50$ & $22.96$ & $23.16$ & $24.97$ & $24.49$ & $23.57$ & $24.36$ & $23.08$ & $21.39$ & $20.21$ & $19.11$ & $23.24$ & $22.50$\\
\hspace{1em}OLS & $-4.38$ & $-4.42$ & $-5.67$ & $-3.25$ & $-3.80$ & $-3.24$ & $-4.80$ & $-3.91$ & $-6.25$ & $-5.13$ & $-4.00$ & $-4.94$ & $-4.11$ & $-4.50$\\
\hspace{1em}WLS$_v$ & $4.45$ & $4.97$ & $5.54$ & $6.53$ & $7.21$ & $7.24$ & $6.61$ & $6.82$ & $5.99$ & $5.36$ & $5.25$ & $4.90$ & $6.02$ & $5.90$\\
\hspace{1em}MinT(Shrink) & $0.22$ & $-0.83$ & $0.39$ & $1.72$ & $1.62$ & $1.41$ & $0.94$ & $0.67$ & $-0.27$ & $-0.18$ & $-0.20$ & $-0.20$ & $0.77$ & $0.42$\\
\hspace{1em}EMinTU & $541.01$ & $474.48$ & $448.10$ & $392.24$ & $335.27$ & $353.31$ & $350.76$ & $322.01$ & $304.31$ & $295.59$ & $271.75$ & $294.01$ & $422.14$ & $359.60$\\
\hspace{1em}IComb$[\alpha = 1, z_{x, y} = 1, c = 1]$ & $-4.95$ & $-3.80$ & $-0.63$ & $-1.91$ & $-2.51$ & $-1.52$ & $-2.78$ & $-4.74$ & $-3.71$ & $-3.93$ & $-6.02$ & $-5.14$ & $-2.53$ & $-3.54$\\
\hspace{1em}IComb$[\alpha = 1, z_{x, y} = 1, c = 0]$ & $-6.50$ & $-5.21$ & $-2.18$ & $-3.01$ & $-3.09$ & $-2.95$ & $-3.31$ & $-3.65$ & $-3.45$ & $-4.05$ & $-5.91$ & $-4.92$ & $-3.79$ & $-4.04$\\
\hspace{1em}IComb$[\alpha = 1, z_{x} = 1, c = 1]$ & $-6.78$ & $-6.48$ & $-2.16$ & $-2.02$ & $-4.64$ & $-5.39$ & $-8.23$ & $-8.22$ & $-7.23$ & $-6.88$ & $-7.58$ & $-6.23$ & $-4.56$ & $-6.07$\\
\hspace{1em}IComb$[\alpha = 1, z_{x} = 1, c = 0]$ & $-8.09$ & $-7.18$ & $-2.76$ & $-3.27$ & $-5.55$ & $-6.67$ & $-8.88$ & $-8.45$ & $-7.86$ & $-7.43$ & $-8.37$ & $-6.45$ & $-5.57$ & $-6.81$\\
\hspace{1em}IComb$[\alpha = 1, z_{x, y} = 0, c = 1]$ & $-8.09$ & $-11.12$ & $-8.88$ & $-7.56$ & $-8.65$ & $-9.30$ & $-12.33$ & $-11.53$ & $-11.44$ & $-11.69$ & $-10.47$ & $-9.99$ & $-8.93$ & $-10.15$\\
\hspace{1em}IComb$[\alpha = 1, z_{x, y} = 0, c = 0]$ & $-9.40$ & $-11.84$ & $-10.94$ & $-7.26$ & $-10.19$ & $-10.58$ & $-13.85$ & $-13.50$ & $-12.95$ & $-14.16$ & $-12.73$ & $-12.66$ & $-10.03$ & $-11.78$\\
\hspace{1em}IComb$[\alpha = 0, z_{x, y} = 1, c = 1]$ & $-6.49$ & $-4.58$ & $-3.58$ & $-3.73$ & $-3.60$ & $-4.02$ & $-7.22$ & $-7.38$ & $-8.89$ & $-10.04$ & $-11.47$ & $-10.55$ & $-4.32$ & $-7.00$\\
\hspace{1em}IComb$[\alpha = 0, z_{x, y} = 1, c = 0]$ & $-7.78$ & $-6.07$ & $-4.68$ & $-4.58$ & $-4.61$ & $-4.40$ & $-7.29$ & $-7.40$ & $-8.23$ & $-8.95$ & $-10.66$ & $-9.65$ & $-5.32$ & $-7.16$\\
\hspace{1em}IComb$[\alpha = 0, z_{x} = 1, c = 1]$ & $-6.35$ & $-4.48$ & $-3.57$ & $-3.68$ & $-3.59$ & $-4.03$ & $-7.30$ & $-7.36$ & $-8.98$ & $-10.10$ & $-11.43$ & $-10.52$ & $-4.26$ & $-6.99$\\
\hspace{1em}IComb$[\alpha = 0, z_{x} = 1, c = 0]$ & $-7.50$ & $-6.16$ & $-4.81$ & $-4.60$ & $-4.37$ & $-4.49$ & $-7.60$ & $-7.40$ & $-8.23$ & $-8.85$ & $-10.63$ & $-9.75$ & $-5.30$ & $-7.17$\\
\hspace{1em}IComb$[\alpha = 0, z_{x, y} = 0, c = 1]$ & $-9.42$ & $-11.25$ & $-10.40$ & $-9.02$ & $-10.22$ & $-10.38$ & $-12.63$ & $-13.13$ & $-14.19$ & $-13.14$ & $-11.82$ & $-11.29$ & $-10.11$ & $-11.48$\\
\hspace{1em}IComb$[\alpha = 0, z_{x, y} = 0, c = 0]$ & $\pmb{-10.20}$ & $\pmb{-12.28}$ & $\pmb{-11.76}$ & $\pmb{-9.77}$ & $\pmb{-11.17}$ & $\pmb{-11.62}$ & $\pmb{-14.00}$ & $\pmb{-14.28}$ & $\pmb{-15.72}$ & $\pmb{-14.98}$ & $\pmb{-13.35}$ & $\pmb{-13.11}$ & $\pmb{-11.13}$ & $\pmb{-12.79}$\\
\cmidrule{2-15}
\multicolumn{15}{l}{\it Zones}\\
\hspace{1em}BU & $4.33$ & $4.49$ & $3.45$ & $3.79$ & $4.77$ & $3.51$ & $4.67$ & $5.34$ & $4.55$ & $4.57$ & $2.23$ & $2.21$ & $4.05$ & $3.98$\\
\hspace{1em}OLS & $-6.32$ & $-6.51$ & $-7.73$ & $-6.68$ & $-6.93$ & $-7.24$ & $-6.84$ & $-6.34$ & $-7.34$ & $-6.63$ & $-6.45$ & $-6.66$ & $-6.91$ & $-6.80$\\
\hspace{1em}WLS$_v$ & $-3.80$ & $-4.12$ & $-5.09$ & $-4.53$ & $-4.32$ & $-4.76$ & $-3.89$ & $-3.67$ & $-4.07$ & $-3.77$ & $-4.53$ & $-4.37$ & $-4.44$ & $-4.24$\\
\hspace{1em}MinT(Shrink) & $-6.41$ & $-7.09$ & $-8.00$ & $-7.26$ & $-7.25$ & $-7.77$ & $-6.86$ & $-6.77$ & $-7.21$ & $-6.70$ & $-7.49$ & $-7.10$ & $-7.30$ & $-7.16$\\
\hspace{1em}EMinTU & $539.68$ & $461.84$ & $454.25$ & $391.82$ & $356.57$ & $354.69$ & $356.60$ & $344.48$ & $336.82$ & $325.06$ & $345.24$ & $317.08$ & $425.20$ & $380.05$\\
\hspace{1em}IComb$[\alpha = 1, z_{x, y} = 1, c = 1]$ & $-12.08$ & $-11.77$ & $-11.25$ & $-12.12$ & $-11.95$ & $-11.95$ & $-10.57$ & $-10.72$ & $-10.54$ & $-9.56$ & $-10.99$ & $-9.86$ & $-11.85$ & $-11.09$\\
\hspace{1em}IComb$[\alpha = 1, z_{x, y} = 1, c = 0]$ & $-13.30$ & $-13.02$ & $-12.45$ & $-13.33$ & $-13.10$ & $-13.40$ & $-11.48$ & $-11.12$ & $-11.54$ & $-10.51$ & $-12.00$ & $-10.78$ & $-13.10$ & $-12.14$\\
\hspace{1em}IComb$[\alpha = 1, z_{x} = 1, c = 1]$ & $-12.18$ & $-12.12$ & $-11.36$ & $-11.03$ & $-11.94$ & $-12.51$ & $-12.30$ & $-11.44$ & $-11.28$ & $-10.24$ & $-10.78$ & $-9.57$ & $-11.86$ & $-11.38$\\
\hspace{1em}IComb$[\alpha = 1, z_{x} = 1, c = 0]$ & $-13.42$ & $-12.74$ & $-12.05$ & $-12.27$ & $-12.90$ & $-13.92$ & $-13.49$ & $-12.58$ & $-12.48$ & $-11.20$ & $-11.82$ & $-10.35$ & $-12.88$ & $-12.41$\\
\hspace{1em}IComb$[\alpha = 1, z_{x, y} = 0, c = 1]$ & $-10.86$ & $-13.52$ & $-13.06$ & $-11.99$ & $-12.93$ & $-12.91$ & $-13.57$ & $-12.71$ & $-12.64$ & $-11.75$ & $-11.30$ & $-10.02$ & $-12.55$ & $-12.25$\\
\hspace{1em}IComb$[\alpha = 1, z_{x, y} = 0, c = 0]$ & $-12.76$ & $-14.64$ & $-14.86$ & $-12.91$ & $-15.44$ & $-14.91$ & $-15.10$ & $-15.16$ & $-14.69$ & $-15.16$ & $-14.55$ & $-13.03$ & $-14.26$ & $-14.44$\\
\hspace{1em}IComb$[\alpha = 0, z_{x, y} = 1, c = 1]$ & $-12.79$ & $-12.63$ & $-12.64$ & $-13.59$ & $-13.09$ & $-13.66$ & $-13.24$ & $-12.53$ & $-14.00$ & $-13.77$ & $-14.99$ & $-13.75$ & $-13.07$ & $-13.41$\\
\hspace{1em}IComb$[\alpha = 0, z_{x, y} = 1, c = 0]$ & $-13.44$ & $-13.31$ & $-13.21$ & $-14.10$ & $-13.72$ & $-13.96$ & $-13.38$ & $-12.70$ & $-14.10$ & $-13.67$ & $-15.35$ & $-14.07$ & $-13.63$ & $-13.76$\\
\hspace{1em}IComb$[\alpha = 0, z_{x} = 1, c = 1]$ & $-12.74$ & $-12.59$ & $-12.62$ & $-13.56$ & $-13.10$ & $-13.67$ & $-13.30$ & $-12.52$ & $-14.05$ & $-13.79$ & $-14.99$ & $-13.75$ & $-13.05$ & $-13.41$\\
\hspace{1em}IComb$[\alpha = 0, z_{x} = 1, c = 0]$ & $-13.26$ & $-13.34$ & $-13.26$ & $-14.11$ & $-13.65$ & $-14.00$ & $-13.49$ & $-12.72$ & $-14.11$ & $-13.63$ & $-15.29$ & $-14.07$ & $-13.61$ & $-13.76$\\
\hspace{1em}IComb$[\alpha = 0, z_{x, y} = 0, c = 1]$ & $-14.28$ & $-15.95$ & $-15.91$ & $-15.30$ & $-15.52$ & $-15.44$ & $-14.97$ & $-15.51$ & $-15.97$ & $-14.49$ & $-14.62$ & $-13.03$ & $-15.40$ & $-15.07$\\
\hspace{1em}IComb$[\alpha = 0, z_{x, y} = 0, c = 0]$ & $\pmb{-15.28}$ & $\pmb{-16.81}$ & $\pmb{-17.24}$ & $\pmb{-16.29}$ & $\pmb{-16.65}$ & $\pmb{-16.52}$ & $\pmb{-16.41}$ & $\pmb{-16.97}$ & $\pmb{-17.80}$ & $\pmb{-16.42}$ & $\pmb{-16.23}$ & $\pmb{-15.01}$ & $\pmb{-16.47}$ & $\pmb{-16.47}$\\
\cmidrule{2-15}
\multicolumn{15}{l}{\it Regions}\\
\hspace{1em}BU & $0.00$ & $0.00$ & $0.00$ & $0.00$ & $0.00$ & $0.00$ & $0.00$ & $0.00$ & $0.00$ & $0.00$ & $0.00$ & $0.00$ & $0.00$ & $0.00$\\
\hspace{1em}OLS & $-4.32$ & $-4.84$ & $-5.13$ & $-4.71$ & $-5.16$ & $-4.73$ & $-4.99$ & $-5.08$ & $-5.32$ & $-4.88$ & $-3.81$ & $-4.07$ & $-4.82$ & $-4.75$\\
\hspace{1em}WLS$_v$ & $-3.49$ & $-4.03$ & $-4.13$ & $-4.00$ & $-4.35$ & $-4.12$ & $-4.13$ & $-4.52$ & $-4.24$ & $-4.10$ & $-3.47$ & $-3.36$ & $-4.02$ & $-4.00$\\
\hspace{1em}MinT(Shrink) & $-4.74$ & $-5.73$ & $-5.93$ & $-5.84$ & $-6.19$ & $-5.89$ & $-5.90$ & $-6.47$ & $-6.04$ & $-5.91$ & $-5.20$ & $-4.95$ & $-5.72$ & $-5.73$\\
\hspace{1em}EMinTU & $560.81$ & $512.11$ & $462.76$ & $412.31$ & $396.78$ & $380.79$ & $375.80$ & $377.99$ & $364.11$ & $363.97$ & $386.29$ & $359.63$ & $453.55$ & $411.47$\\
\hspace{1em}IComb$[\alpha = 1, z_{x, y} = 1, c = 1]$ & $-12.94$ & $-13.36$ & $-12.60$ & $-13.06$ & $-13.28$ & $-12.85$ & $-12.40$ & $-12.48$ & $-12.34$ & $-10.95$ & $-11.08$ & $-10.40$ & $-13.02$ & $-12.29$\\
\hspace{1em}IComb$[\alpha = 1, z_{x, y} = 1, c = 0]$ & $-14.03$ & $-14.53$ & $-13.49$ & $-14.24$ & $-14.46$ & $-14.03$ & $-13.42$ & $-13.35$ & $-13.69$ & $-12.41$ & $-12.85$ & $-12.09$ & $-14.13$ & $-13.53$\\
\hspace{1em}IComb$[\alpha = 1, z_{x} = 1, c = 1]$ & $-13.27$ & $-13.85$ & $-13.01$ & $-12.62$ & $-12.97$ & $-13.06$ & $-13.05$ & $-12.57$ & $-12.52$ & $-11.16$ & $-10.39$ & $-10.07$ & $-13.13$ & $-12.36$\\
\hspace{1em}IComb$[\alpha = 1, z_{x} = 1, c = 0]$ & $-14.35$ & $-14.55$ & $-13.79$ & $-13.96$ & $-14.28$ & $-14.45$ & $-14.26$ & $-14.15$ & $-14.07$ & $-12.63$ & $-12.13$ & $-11.61$ & $-14.23$ & $-13.67$\\
\hspace{1em}IComb$[\alpha = 1, z_{x, y} = 0, c = 1]$ & $-10.47$ & $-12.83$ & $-12.19$ & $-11.86$ & $-12.26$ & $-11.86$ & $-12.38$ & $-11.98$ & $-11.39$ & $-10.06$ & $-8.67$ & $-7.67$ & $-11.91$ & $-11.11$\\
\hspace{1em}IComb$[\alpha = 1, z_{x, y} = 0, c = 0]$ & $-11.73$ & $-13.54$ & $-13.76$ & $-13.09$ & $-14.70$ & $-13.37$ & $-14.47$ & $-14.54$ & $-14.05$ & $-13.29$ & $-12.01$ & $-11.04$ & $-13.37$ & $-13.29$\\
\hspace{1em}IComb$[\alpha = 0, z_{x, y} = 1, c = 1]$ & $-13.71$ & $-14.02$ & $-13.59$ & $-14.21$ & $-14.12$ & $-14.27$ & $-14.08$ & $-13.95$ & $-14.67$ & $-13.43$ & $-13.96$ & $-13.11$ & $-13.99$ & $-13.92$\\
\hspace{1em}IComb$[\alpha = 0, z_{x, y} = 1, c = 0]$ & $-14.16$ & $-14.53$ & $-13.90$ & $-14.75$ & $-14.73$ & $-14.76$ & $-14.45$ & $-14.37$ & $-15.11$ & $-13.73$ & $\pmb{-14.69}$ & $\pmb{-13.83}$ & $-14.48$ & $-14.42$\\
\hspace{1em}IComb$[\alpha = 0, z_{x} = 1, c = 1]$ & $-13.67$ & $-13.99$ & $-13.57$ & $-14.19$ & $-14.10$ & $-14.27$ & $-14.12$ & $-13.96$ & $-14.70$ & $-13.41$ & $-13.95$ & $-13.11$ & $-13.97$ & $-13.92$\\
\hspace{1em}IComb$[\alpha = 0, z_{x} = 1, c = 0]$ & $-14.07$ & $-14.54$ & $-13.96$ & $-14.75$ & $-14.67$ & $-14.81$ & $-14.50$ & $-14.38$ & $-15.12$ & $-13.73$ & $-14.64$ & $-13.81$ & $-14.47$ & $-14.41$\\
\hspace{1em}IComb$[\alpha = 0, z_{x, y} = 0, c = 1]$ & $-14.09$ & $-15.24$ & $-15.10$ & $-15.21$ & $-15.05$ & $-14.78$ & $-14.38$ & $-14.71$ & $-14.92$ & $-13.19$ & $-12.21$ & $-11.10$ & $-14.91$ & $-14.14$\\
\hspace{1em}IComb$[\alpha = 0, z_{x, y} = 0, c = 0]$ & $\pmb{-15.22}$ & $\pmb{-16.37}$ & $\pmb{-16.32}$ & $\pmb{-16.30}$ & $\pmb{-16.21}$ & $\pmb{-15.78}$ & $\pmb{-15.89}$ & $\pmb{-16.49}$ & $\pmb{-16.99}$ & $\pmb{-15.42}$ & $-14.45$ & $-13.57$ & $\pmb{-16.03}$ & $\pmb{-15.74}$\\
\cmidrule{2-15}
\multicolumn{15}{l}{\it Average}\\
\hspace{1em}BU & $30.09$ & $34.51$ & $35.05$ & $31.32$ & $33.52$ & $33.44$ & $36.28$ & $35.54$ & $40.35$ & $35.47$ & $31.63$ & $32.84$ & $32.99$ & $34.19$\\
\hspace{1em}OLS & $-2.71$ & $-2.83$ & $-3.04$ & $-2.42$ & $-2.56$ & $-2.53$ & $-2.37$ & $-2.33$ & $-2.26$ & $-2.12$ & $-2.04$ & $-2.03$ & $-2.68$ & $-2.41$\\
\hspace{1em}WLS$_v$ & $10.55$ & $12.36$ & $14.22$ & $11.52$ & $12.82$ & $12.53$ & $15.21$ & $14.21$ & $18.38$ & $15.07$ & $12.55$ & $14.07$ & $12.34$ & $13.68$\\
\hspace{1em}MinT(Shrink) & $4.58$ & $4.37$ & $7.13$ & $4.83$ & $5.17$ & $4.46$ & $7.29$ & $5.72$ & $9.28$ & $6.99$ & $4.75$ & $6.60$ & $5.08$ & $5.96$\\
\hspace{1em}EMinTU & $462.60$ & $455.04$ & $432.07$ & $362.15$ & $320.65$ & $331.74$ & $332.91$ & $322.22$ & $306.60$ & $303.26$ & $235.94$ & $289.18$ & $391.53$ & $341.15$\\
\hspace{1em}IComb$[\alpha = 1, z_{x, y} = 1, c = 1]$ & $-2.44$ & $-0.93$ & $3.39$ & $-1.85$ & $-1.21$ & $-1.19$ & $0.85$ & $-3.39$ & $2.33$ & $-0.11$ & $-5.25$ & $-2.28$ & $-0.72$ & $-1.08$\\
\hspace{1em}IComb$[\alpha = 1, z_{x, y} = 1, c = 0]$ & $-3.90$ & $-2.16$ & $1.96$ & $-2.73$ & $-1.54$ & $-1.97$ & $1.13$ & $-1.02$ & $3.02$ & $0.51$ & $-4.23$ & $-1.16$ & $-1.72$ & $-1.00$\\
\hspace{1em}IComb$[\alpha = 1, z_{x} = 1, c = 1]$ & $-6.63$ & $-6.18$ & $1.48$ & $-3.21$ & $-3.72$ & $-6.29$ & $-6.33$ & $-7.95$ & $-2.76$ & $-4.05$ & $-7.25$ & $-3.40$ & $-4.09$ & $-4.73$\\
\hspace{1em}IComb$[\alpha = 1, z_{x} = 1, c = 0]$ & $-6.74$ & $-5.77$ & $1.86$ & $-3.54$ & $-4.53$ & $-7.04$ & $-6.02$ & $-7.23$ & $-3.10$ & $-3.47$ & $-7.35$ & $-2.93$ & $-4.31$ & $-4.68$\\
\hspace{1em}IComb$[\alpha = 1, z_{x, y} = 0, c = 1]$ & $-5.63$ & $-10.18$ & $-5.20$ & $-7.23$ & $-7.95$ & $-9.47$ & $-10.21$ & $-9.79$ & $-6.41$ & $-7.57$ & $-8.46$ & $-5.79$ & $-7.64$ & $-7.83$\\
\hspace{1em}IComb$[\alpha = 1, z_{x, y} = 0, c = 0]$ & $-7.96$ & $-10.16$ & $-7.07$ & $-6.58$ & $-9.96$ & $-11.49$ & $\pmb{-12.63}$ & $-12.33$ & $-8.78$ & $-10.82$ & $-11.42$ & $-8.83$ & $-8.90$ & $-9.90$\\
\hspace{1em}IComb$[\alpha = 0, z_{x, y} = 1, c = 1]$ & $-4.36$ & $-2.70$ & $-0.24$ & $-3.30$ & $-2.89$ & $-4.49$ & $-5.33$ & $-6.69$ & $-3.94$ & $-6.96$ & $-11.38$ & $-8.64$ & $-3.02$ & $-5.28$\\
\hspace{1em}IComb$[\alpha = 0, z_{x, y} = 1, c = 0]$ & $-5.07$ & $-3.60$ & $-0.25$ & $-3.19$ & $-2.53$ & $-3.55$ & $-4.12$ & $-5.27$ & $-1.80$ & $-4.17$ & $-8.97$ & $-5.74$ & $-3.03$ & $-4.13$\\
\hspace{1em}IComb$[\alpha = 0, z_{x} = 1, c = 1]$ & $-4.24$ & $-2.57$ & $-0.21$ & $-3.27$ & $-2.90$ & $-4.47$ & $-5.41$ & $-6.69$ & $-4.03$ & $-7.07$ & $-11.33$ & $-8.60$ & $-2.96$ & $-5.27$\\
\hspace{1em}IComb$[\alpha = 0, z_{x} = 1, c = 0]$ & $-4.97$ & $-3.70$ & $-0.40$ & $-3.20$ & $-2.20$ & $-3.74$ & $-4.49$ & $-5.14$ & $-1.94$ & $-4.15$ & $-8.91$ & $-5.85$ & $-3.03$ & $-4.16$\\
\hspace{1em}IComb$[\alpha = 0, z_{x, y} = 0, c = 1]$ & $-8.06$ & $-9.89$ & $-7.72$ & $-9.13$ & $-10.15$ & $-11.26$ & $-10.52$ & $-12.30$ & $-10.57$ & $-10.16$ & $-10.73$ & $-8.22$ & $-9.41$ & $-9.93$\\
\hspace{1em}IComb$[\alpha = 0, z_{x, y} = 0, c = 0]$ & $\pmb{-8.99}$ & $\pmb{-10.85}$ & $\pmb{-8.52}$ & $\pmb{-9.84}$ & $\pmb{-11.15}$ & $\pmb{-12.45}$ & $-12.11$ & $\pmb{-13.61}$ & $\pmb{-12.69}$ & $\pmb{-12.45}$ & $\pmb{-12.88}$ & $\pmb{-10.52}$ & $\pmb{-10.34}$ & $\pmb{-11.41}$\\
\end{longtable}
\end{landscape}
\end{footnotesize}

\subsection{ETS forecasts}

\begin{landscape}
\begin{footnotesize}
\begin{longtable}[h]{lrrrrrrrrrrrrrr}
\caption{\mbox{PRIAL for forecast reconciliation methods using ETS base forecasts.}}\\
\toprule
& \multicolumn{14}{c}{Forecast horizon $(h)$}\\
\cmidrule{2-15}
  & {\it 1} & {\it 2} & {\it 3} & {\it 4} & {\it 5} & {\it 6} & {\it 7} & {\it 8} & {\it 9} & {\it 10} & {\it 11} & {\it 12} & {\it 1--6} & {\it 1--12}\\
 \cmidrule{2-15}
\endfirsthead
\caption{(continued from previous page)}\\
\toprule
& \multicolumn{14}{c}{Forecast horizon $(h)$}\\
\cmidrule{2-15}
 & {\it 1} & {\it 2} & {\it 3} & {\it 4} & {\it 5} & {\it 6} & {\it 7} & {\it 8} & {\it 9} & {\it 10} & {\it 11} & {\it 12} & {\it 1--6} & {\it 1--12}\\
 \cmidrule{2-15}
\endhead
\bottomrule
\endfoot
\multicolumn{15}{l}{\it Australia}\\
\hspace{1em}BU & $37.41$ & $28.81$ & $27.54$ & $26.57$ & $29.33$ & $32.47$ & $32.88$ & $39.52$ & $38.26$ & $37.70$ & $38.09$ & $38.99$ & $30.27$ & $34.41$\\
\hspace{1em}OLS & $0.07$ & $-0.31$ & $-0.94$ & $-1.31$ & $-1.29$ & $-0.77$ & $-0.32$ & $0.11$ & $-0.26$ & $-0.23$ & $-0.08$ & $-0.15$ & $-0.78$ & $-0.43$\\
\hspace{1em}WLS$_v$ & $17.65$ & $13.14$ & $11.32$ & $9.59$ & $10.93$ & $13.87$ & $15.76$ & $19.35$ & $18.24$ & $18.00$ & $18.45$ & $18.76$ & $12.66$ & $15.73$\\
\hspace{1em}MinT(Shrink) & $12.91$ & $8.79$ & $6.90$ & $4.67$ & $6.25$ & $8.65$ & $10.55$ & $13.30$ & $12.06$ & $12.19$ & $13.01$ & $12.54$ & $7.93$ & $10.38$\\
\hspace{1em}EMinTU & $792.26$ & $1174.87$ & $1648.94$ & $1785.04$ & $2195.41$ & $2379.61$ & $2909.64$ & $2936.09$ & $2825.55$ & $3066.33$ & $2922.22$ & $3074.02$ & $1693.24$ & $2403.10$\\
\hspace{1em}IComb$[\alpha = 1, z_{x, y} = 1, c = 1]$ & $-5.94$ & $-6.75$ & $-5.26$ & $-8.23$ & $-6.45$ & $-8.53$ & $-7.11$ & $-7.94$ & $-6.75$ & $-4.70$ & $-3.34$ & $-2.22$ & $-6.90$ & $-5.93$\\
\hspace{1em}IComb$[\alpha = 1, z_{x, y} = 1, c = 0]$ & $\pmb{-6.99}$ & $\pmb{-8.37}$ & $-4.91$ & $\pmb{-9.59}$ & $-8.86$ & $\pmb{-11.25}$ & $-10.08$ & $\pmb{-10.42}$ & $-9.21$ & $-7.99$ & $-6.99$ & $-6.61$ & $\pmb{-8.40}$ & $\pmb{-8.41}$\\
\hspace{1em}IComb$[\alpha = 1, z_{x} = 1, c = 1]$ & $-2.51$ & $-0.36$ & $-3.97$ & $-7.09$ & $-8.22$ & $-9.85$ & $\pmb{-10.15}$ & $-8.81$ & $-11.52$ & $-6.96$ & $-8.43$ & $-8.96$ & $-5.52$ & $-7.52$\\
\hspace{1em}IComb$[\alpha = 1, z_{x} = 1, c = 0]$ & $-2.05$ & $-0.26$ & $-3.16$ & $-6.05$ & $-7.06$ & $-9.92$ & $-10.11$ & $-9.54$ & $\pmb{-11.98}$ & $-7.11$ & $-8.72$ & $\pmb{-9.26}$ & $-4.93$ & $-7.43$\\
\hspace{1em}IComb$[\alpha = 1, z_{x, y} = 0, c = 1]$ & $-3.92$ & $-2.69$ & $-3.88$ & $-6.69$ & $-7.89$ & $-8.69$ & $-8.16$ & $-7.06$ & $-8.97$ & $-7.38$ & $-7.58$ & $-7.82$ & $-5.75$ & $-6.91$\\
\hspace{1em}IComb$[\alpha = 1, z_{x, y} = 0, c = 0]$ & $-4.91$ & $-3.02$ & $\pmb{-5.71}$ & $-6.97$ & $\pmb{-8.92}$ & $-9.56$ & $-9.24$ & $-8.34$ & $-9.60$ & $\pmb{-8.81}$ & $-8.71$ & $-8.58$ & $-6.64$ & $-7.89$\\
\hspace{1em}IComb$[\alpha = 0, z_{x, y} = 1, c = 1]$ & $-1.16$ & $-3.50$ & $-2.93$ & $-5.59$ & $-4.29$ & $-4.73$ & $-4.22$ & $-2.57$ & $-2.84$ & $-1.26$ & $-0.19$ & $-0.41$ & $-3.77$ & $-2.67$\\
\hspace{1em}IComb$[\alpha = 0, z_{x, y} = 1, c = 0]$ & $-0.47$ & $-3.17$ & $-3.04$ & $-5.56$ & $-4.85$ & $-5.02$ & $-4.71$ & $-2.64$ & $-2.59$ & $-1.49$ & $-0.40$ & $-0.60$ & $-3.77$ & $-2.76$\\
\hspace{1em}IComb$[\alpha = 0, z_{x} = 1, c = 1]$ & $-1.35$ & $-3.39$ & $-2.86$ & $-5.72$ & $-4.36$ & $-4.72$ & $-4.21$ & $-2.33$ & $-2.80$ & $-1.27$ & $-0.22$ & $-0.23$ & $-3.80$ & $-2.64$\\
\hspace{1em}IComb$[\alpha = 0, z_{x} = 1, c = 0]$ & $-0.54$ & $-3.32$ & $-2.46$ & $-5.21$ & $-4.42$ & $-4.64$ & $-4.23$ & $-2.07$ & $-2.11$ & $-1.42$ & $-0.41$ & $-0.49$ & $-3.51$ & $-2.49$\\
\hspace{1em}IComb$[\alpha = 0, z_{x, y} = 0, c = 1]$ & $-4.13$ & $-3.17$ & $-4.21$ & $-6.08$ & $-7.64$ & $-7.73$ & $-7.89$ & $-7.68$ & $-9.39$ & $-7.89$ & $-8.15$ & $-8.11$ & $-5.59$ & $-7.04$\\
\hspace{1em}IComb$[\alpha = 0, z_{x, y} = 0, c = 0]$ & $-5.10$ & $-3.68$ & $-5.34$ & $-6.92$ & $-8.53$ & $-8.34$ & $-8.27$ & $-8.50$ & $-9.80$ & $-8.64$ & $\pmb{-9.02}$ & $-8.76$ & $-6.41$ & $-7.77$\\
\cmidrule{2-15}
\multicolumn{15}{l}{\it States}\\
\hspace{1em}BU & $10.34$ & $7.36$ & $8.70$ & $13.20$ & $12.53$ & $11.08$ & $11.41$ & $14.68$ & $13.83$ & $20.13$ & $17.61$ & $19.31$ & $10.55$ & $13.59$\\
\hspace{1em}OLS & $-2.55$ & $-1.78$ & $-0.59$ & $1.22$ & $0.93$ & $-0.24$ & $-0.58$ & $-0.53$ & $-0.19$ & $0.89$ & $0.28$ & $0.67$ & $-0.49$ & $-0.17$\\
\hspace{1em}WLS$_v$ & $1.91$ & $1.03$ & $2.01$ & $5.04$ & $4.49$ & $3.23$ & $4.03$ & $5.67$ & $5.01$ & $8.79$ & $7.34$ & $8.30$ & $2.96$ & $4.89$\\
\hspace{1em}MinT(Shrink) & $-0.26$ & $-0.88$ & $-0.04$ & $2.59$ & $2.18$ & $0.81$ & $1.45$ & $2.55$ & $1.68$ & $5.77$ & $4.55$ & $5.33$ & $0.74$ & $2.27$\\
\hspace{1em}EMinTU & $925.42$ & $1324.17$ & $1683.98$ & $1990.22$ & $2748.09$ & $2804.73$ & $3128.05$ & $3409.19$ & $3535.36$ & $3692.96$ & $3895.42$ & $3789.60$ & $1924.69$ & $2809.33$\\
\hspace{1em}IComb$[\alpha = 1, z_{x, y} = 1, c = 1]$ & $-7.69$ & $-6.91$ & $-4.30$ & $-1.80$ & $-2.68$ & $-5.49$ & $-5.28$ & $-4.03$ & $-4.98$ & $-2.64$ & $-3.37$ & $-1.46$ & $-4.80$ & $-4.15$\\
\hspace{1em}IComb$[\alpha = 1, z_{x, y} = 1, c = 0]$ & $-7.88$ & $\pmb{-7.35}$ & $-3.81$ & $-2.08$ & $-3.60$ & $-7.18$ & $-6.65$ & $-5.45$ & $-5.90$ & $-4.16$ & $-4.39$ & $-3.29$ & $-5.32$ & $-5.11$\\
\hspace{1em}IComb$[\alpha = 1, z_{x} = 1, c = 1]$ & $-5.06$ & $-3.10$ & $-4.11$ & $-1.37$ & $-4.60$ & $-7.54$ & $\pmb{-8.10}$ & $-5.75$ & $-7.13$ & $-3.83$ & $-4.97$ & $-5.75$ & $-4.32$ & $-5.16$\\
\hspace{1em}IComb$[\alpha = 1, z_{x} = 1, c = 0]$ & $-6.03$ & $-3.04$ & $-3.62$ & $-0.53$ & $-4.12$ & $\pmb{-7.93}$ & $-7.84$ & $-5.80$ & $-7.52$ & $-4.95$ & $-5.81$ & $-5.84$ & $-4.24$ & $-5.33$\\
\hspace{1em}IComb$[\alpha = 1, z_{x, y} = 0, c = 1]$ & $-8.03$ & $-5.49$ & $-4.69$ & $-3.11$ & $-4.49$ & $-6.05$ & $-6.54$ & $-6.11$ & $-7.69$ & $-5.46$ & $-6.34$ & $-5.93$ & $-5.31$ & $-5.86$\\
\hspace{1em}IComb$[\alpha = 1, z_{x, y} = 0, c = 0]$ & $-9.23$ & $-6.32$ & $-5.87$ & $-4.02$ & $\pmb{-5.90}$ & $-7.42$ & $-7.49$ & $-7.72$ & $-8.45$ & $\pmb{-6.51}$ & $-6.57$ & $-6.52$ & $-6.46$ & $-6.85$\\
\hspace{1em}IComb$[\alpha = 0, z_{x, y} = 1, c = 1]$ & $-6.74$ & $-6.86$ & $-5.57$ & $-2.34$ & $-2.99$ & $-5.29$ & $-5.70$ & $-3.28$ & $-4.54$ & $-0.78$ & $-1.83$ & $-1.58$ & $-4.96$ & $-3.86$\\
\hspace{1em}IComb$[\alpha = 0, z_{x, y} = 1, c = 0]$ & $-6.98$ & $-7.30$ & $-5.79$ & $-1.94$ & $-3.04$ & $-5.40$ & $-5.99$ & $-3.14$ & $-4.16$ & $-0.90$ & $-1.84$ & $-1.62$ & $-5.07$ & $-3.91$\\
\hspace{1em}IComb$[\alpha = 0, z_{x} = 1, c = 1]$ & $-6.87$ & $-6.79$ & $-5.53$ & $-2.34$ & $-3.02$ & $-5.35$ & $-5.70$ & $-3.12$ & $-4.49$ & $-0.79$ & $-1.85$ & $-1.50$ & $-4.98$ & $-3.85$\\
\hspace{1em}IComb$[\alpha = 0, z_{x} = 1, c = 0]$ & $-6.90$ & $-7.29$ & $-5.60$ & $-1.89$ & $-2.82$ & $-5.12$ & $-5.69$ & $-2.77$ & $-3.86$ & $-0.62$ & $-1.90$ & $-1.65$ & $-4.93$ & $-3.75$\\
\hspace{1em}IComb$[\alpha = 0, z_{x, y} = 0, c = 1]$ & $-8.67$ & $-6.63$ & $-6.00$ & $-3.60$ & $-5.08$ & $-6.65$ & $-7.68$ & $-7.04$ & $-8.49$ & $-5.61$ & $-6.88$ & $-6.67$ & $-6.10$ & $-6.61$\\
\hspace{1em}IComb$[\alpha = 0, z_{x, y} = 0, c = 0]$ & $\pmb{-9.33}$ & $-7.12$ & $\pmb{-6.69}$ & $\pmb{-4.18}$ & $-5.76$ & $-7.26$ & $-8.07$ & $\pmb{-7.88}$ & $\pmb{-8.89}$ & $-6.34$ & $\pmb{-7.47}$ & $\pmb{-6.98}$ & $\pmb{-6.72}$ & $\pmb{-7.19}$\\
\cmidrule{2-15}
\multicolumn{15}{l}{\it Zones}\\
\hspace{1em}BU & $0.94$ & $1.13$ & $1.46$ & $2.41$ & $2.51$ & $1.96$ & $1.11$ & $2.31$ & $2.48$ & $5.73$ & $3.57$ & $3.70$ & $1.74$ & $2.49$\\
\hspace{1em}OLS & $-3.81$ & $-2.43$ & $-2.46$ & $-2.69$ & $-2.82$ & $-2.84$ & $-3.57$ & $-4.14$ & $-3.75$ & $-3.97$ & $-4.28$ & $-4.56$ & $-2.84$ & $-3.47$\\
\hspace{1em}WLS$_v$ & $-2.78$ & $-1.98$ & $-1.86$ & $-1.79$ & $-2.00$ & $-2.25$ & $-2.67$ & $-2.55$ & $-2.56$ & $-1.32$ & $-2.29$ & $-2.37$ & $-2.11$ & $-2.20$\\
\hspace{1em}MinT(Shrink) & $-3.41$ & $-2.64$ & $-2.66$ & $-2.71$ & $-2.88$ & $-3.06$ & $-3.67$ & $-3.78$ & $-3.80$ & $-2.50$ & $-3.42$ & $-3.52$ & $-2.90$ & $-3.18$\\
\hspace{1em}EMinTU & $1144.09$ & $1561.37$ & $2094.00$ & $2396.67$ & $2924.76$ & $3327.26$ & $3653.74$ & $4148.44$ & $4283.72$ & $4669.22$ & $4855.42$ & $4647.70$ & $2251.60$ & $3363.72$\\
\hspace{1em}IComb$[\alpha = 1, z_{x, y} = 1, c = 1]$ & $-10.08$ & $-8.08$ & $-7.54$ & $-6.35$ & $-6.63$ & $-8.68$ & $-9.81$ & $-9.94$ & $-9.09$ & $-8.40$ & $-8.84$ & $-8.57$ & $-7.89$ & $-8.52$\\
\hspace{1em}IComb$[\alpha = 1, z_{x, y} = 1, c = 0]$ & $\pmb{-10.37}$ & $-8.62$ & $-7.56$ & $-6.44$ & $-7.26$ & $\pmb{-9.55}$ & $-10.43$ & $-10.54$ & $-9.36$ & $-9.21$ & $-9.81$ & $-9.72$ & $-8.30$ & $-9.10$\\
\hspace{1em}IComb$[\alpha = 1, z_{x} = 1, c = 1]$ & $-5.86$ & $-3.56$ & $-4.21$ & $-3.39$ & $-4.74$ & $-6.58$ & $-7.61$ & $-8.23$ & $-7.38$ & $-6.20$ & $-6.74$ & $-7.50$ & $-4.74$ & $-6.06$\\
\hspace{1em}IComb$[\alpha = 1, z_{x} = 1, c = 0]$ & $-6.61$ & $-3.53$ & $-4.19$ & $-3.27$ & $-4.92$ & $-6.90$ & $-7.69$ & $-8.26$ & $-7.79$ & $-6.97$ & $-7.38$ & $-8.00$ & $-4.92$ & $-6.36$\\
\hspace{1em}IComb$[\alpha = 1, z_{x, y} = 0, c = 1]$ & $-3.20$ & $-0.53$ & $-0.90$ & $-0.43$ & $-1.43$ & $-2.63$ & $-3.63$ & $-4.72$ & $-4.21$ & $-3.76$ & $-4.52$ & $-4.53$ & $-1.53$ & $-2.94$\\
\hspace{1em}IComb$[\alpha = 1, z_{x, y} = 0, c = 0]$ & $-7.04$ & $-3.87$ & $-4.10$ & $-3.90$ & $-5.34$ & $-6.59$ & $-6.95$ & $-8.28$ & $-7.79$ & $-7.26$ & $-8.19$ & $-8.29$ & $-5.15$ & $-6.53$\\
\hspace{1em}IComb$[\alpha = 0, z_{x, y} = 1, c = 1]$ & $-9.65$ & $-8.21$ & $-8.25$ & $-7.41$ & $-7.74$ & $-9.15$ & $-10.24$ & $-9.67$ & $-9.23$ & $-7.92$ & $-8.84$ & $-9.03$ & $-8.41$ & $-8.79$\\
\hspace{1em}IComb$[\alpha = 0, z_{x, y} = 1, c = 0]$ & $-9.92$ & $-8.62$ & $\pmb{-8.59}$ & $-7.38$ & $\pmb{-8.01}$ & $-9.37$ & $\pmb{-10.56}$ & $-9.72$ & $-9.14$ & $-8.10$ & $-9.12$ & $-9.30$ & $\pmb{-8.65}$ & $-9.00$\\
\hspace{1em}IComb$[\alpha = 0, z_{x} = 1, c = 1]$ & $-9.74$ & $-8.18$ & $-8.25$ & $\pmb{-7.42}$ & $-7.78$ & $-9.17$ & $-10.27$ & $-9.61$ & $-9.20$ & $-7.88$ & $-8.85$ & $-8.97$ & $-8.42$ & $-8.79$\\
\hspace{1em}IComb$[\alpha = 0, z_{x} = 1, c = 0]$ & $-9.90$ & $\pmb{-8.71}$ & $-8.49$ & $-7.32$ & $-7.88$ & $-9.26$ & $-10.39$ & $-9.51$ & $-8.97$ & $-7.89$ & $-9.10$ & $-9.32$ & $-8.59$ & $-8.91$\\
\hspace{1em}IComb$[\alpha = 0, z_{x, y} = 0, c = 1]$ & $-7.73$ & $-5.48$ & $-5.69$ & $-5.42$ & $-6.17$ & $-7.09$ & $-8.21$ & $-9.04$ & $-8.65$ & $-7.94$ & $-8.98$ & $-8.99$ & $-6.27$ & $-7.50$\\
\hspace{1em}IComb$[\alpha = 0, z_{x, y} = 0, c = 0]$ & $-9.32$ & $-7.11$ & $-7.24$ & $-6.86$ & $-7.90$ & $-8.81$ & $-9.80$ & $\pmb{-10.86}$ & $\pmb{-10.22}$ & $\pmb{-9.52}$ & $\pmb{-10.70}$ & $\pmb{-10.60}$ & $-7.88$ & $\pmb{-9.13}$\\
\cmidrule{2-15}
\multicolumn{15}{l}{\it Regions}\\
\hspace{1em}BU & $0.00$ & $0.00$ & $0.00$ & $0.00$ & $0.00$ & $0.00$ & $0.00$ & $0.00$ & $0.00$ & $0.00$ & $0.00$ & $0.00$ & $0.00$ & $0.00$\\
\hspace{1em}OLS & $-2.17$ & $-1.96$ & $-2.06$ & $-2.69$ & $-2.61$ & $-2.42$ & $-2.38$ & $-3.20$ & $-2.99$ & $-4.50$ & $-3.71$ & $-3.77$ & $-2.32$ & $-2.90$\\
\hspace{1em}WLS$_v$ & $-1.92$ & $-1.73$ & $-1.84$ & $-2.21$ & $-2.39$ & $-2.26$ & $-2.13$ & $-2.84$ & $-2.80$ & $-3.94$ & $-3.50$ & $-3.23$ & $-2.06$ & $-2.59$\\
\hspace{1em}MinT(Shrink) & $-2.41$ & $-2.17$ & $-2.39$ & $-2.82$ & $-3.02$ & $-2.89$ & $-2.79$ & $-3.59$ & $-3.43$ & $-4.66$ & $-4.19$ & $-3.86$ & $-2.62$ & $-3.21$\\
\hspace{1em}EMinTU & $1224.69$ & $1623.45$ & $2007.34$ & $2460.61$ & $3041.47$ & $3539.59$ & $3838.63$ & $4348.14$ & $4545.91$ & $4893.62$ & $5186.61$ & $4888.86$ & $2325.00$ & $3516.26$\\
\hspace{1em}IComb$[\alpha = 1, z_{x, y} = 1, c = 1]$ & $-9.20$ & $-7.59$ & $-7.82$ & $-7.71$ & $-7.52$ & $-8.36$ & $-8.21$ & $-9.61$ & $-8.54$ & $-9.84$ & $-8.35$ & $-8.66$ & $-8.04$ & $-8.47$\\
\hspace{1em}IComb$[\alpha = 1, z_{x, y} = 1, c = 0]$ & $-9.33$ & $-8.01$ & $-7.89$ & $-7.87$ & $-7.91$ & $-8.97$ & $-9.03$ & $-10.25$ & $-8.79$ & $-10.40$ & $-9.06$ & $-9.37$ & $-8.34$ & $-8.93$\\
\hspace{1em}IComb$[\alpha = 1, z_{x} = 1, c = 1]$ & $-5.49$ & $-3.44$ & $-3.99$ & $-4.32$ & $-4.68$ & $-5.27$ & $-5.50$ & $-6.81$ & $-5.94$ & $-6.98$ & $-5.31$ & $-5.48$ & $-4.54$ & $-5.30$\\
\hspace{1em}IComb$[\alpha = 1, z_{x} = 1, c = 0]$ & $-6.00$ & $-3.64$ & $-4.23$ & $-4.41$ & $-4.96$ & $-5.55$ & $-5.73$ & $-6.64$ & $-6.30$ & $-7.73$ & $-5.88$ & $-6.07$ & $-4.81$ & $-5.63$\\
\hspace{1em}IComb$[\alpha = 1, z_{x, y} = 0, c = 1]$ & $-1.38$ & $0.19$ & $-0.16$ & $-0.03$ & $-0.56$ & $-1.11$ & $-1.17$ & $-2.76$ & $-1.80$ & $-3.15$ & $-1.30$ & $-1.10$ & $-0.52$ & $-1.22$\\
\hspace{1em}IComb$[\alpha = 1, z_{x, y} = 0, c = 0]$ & $-5.30$ & $-3.39$ & $-4.06$ & $-4.13$ & $-4.88$ & $-5.83$ & $-5.62$ & $-7.26$ & $-6.62$ & $-8.08$ & $-6.61$ & $-6.57$ & $-4.61$ & $-5.74$\\
\hspace{1em}IComb$[\alpha = 0, z_{x, y} = 1, c = 1]$ & $-9.17$ & $-7.86$ & $-8.25$ & $-8.26$ & $-8.43$ & $-8.99$ & $-9.15$ & $-9.71$ & $-8.97$ & $-10.00$ & $-8.73$ & $-9.09$ & $-8.50$ & $-8.90$\\
\hspace{1em}IComb$[\alpha = 0, z_{x, y} = 1, c = 0]$ & $\pmb{-9.51}$ & $-8.28$ & $\pmb{-8.65}$ & $\pmb{-8.37}$ & $\pmb{-8.73}$ & $\pmb{-9.30}$ & $\pmb{-9.58}$ & $-9.91$ & $-9.10$ & $-10.26$ & $-9.04$ & $-9.44$ & $\pmb{-8.81}$ & $\pmb{-9.20}$\\
\hspace{1em}IComb$[\alpha = 0, z_{x} = 1, c = 1]$ & $-9.20$ & $-7.83$ & $-8.24$ & $-8.25$ & $-8.47$ & $-8.99$ & $-9.16$ & $-9.67$ & $-8.97$ & $-9.97$ & $-8.74$ & $-9.05$ & $-8.50$ & $-8.89$\\
\hspace{1em}IComb$[\alpha = 0, z_{x} = 1, c = 0]$ & $-9.46$ & $\pmb{-8.30}$ & $-8.56$ & $-8.29$ & $-8.60$ & $-9.19$ & $-9.45$ & $-9.78$ & $-8.99$ & $-10.11$ & $-9.01$ & $-9.45$ & $-8.74$ & $-9.11$\\
\hspace{1em}IComb$[\alpha = 0, z_{x, y} = 0, c = 1]$ & $-7.04$ & $-5.47$ & $-5.85$ & $-6.15$ & $-6.60$ & $-6.92$ & $-6.88$ & $-8.13$ & $-7.48$ & $-8.74$ & $-7.53$ & $-7.55$ & $-6.35$ & $-7.06$\\
\hspace{1em}IComb$[\alpha = 0, z_{x, y} = 0, c = 0]$ & $-8.42$ & $-7.16$ & $-7.53$ & $-7.80$ & $-8.47$ & $-8.84$ & $-8.80$ & $\pmb{-10.32}$ & $\pmb{-9.54}$ & $\pmb{-10.77}$ & $\pmb{-9.68}$ & $\pmb{-9.69}$ & $-8.05$ & $-8.96$\\
\cmidrule{2-15}
\multicolumn{15}{l}{\it Average}\\
\hspace{1em}BU & $19.86$ & $15.37$ & $15.41$ & $16.41$ & $17.58$ & $18.68$ & $19.04$ & $23.38$ & $22.94$ & $24.93$ & $24.32$ & $25.41$ & $17.23$ & $20.60$\\
\hspace{1em}OLS & $-1.49$ & $-1.23$ & $-1.24$ & $-1.12$ & $-1.19$ & $-1.17$ & $-1.12$ & $-1.08$ & $-1.08$ & $-1.01$ & $-1.01$ & $-0.98$ & $-1.24$ & $-1.13$\\
\hspace{1em}WLS$_v$ & $7.87$ & $5.82$ & $5.34$ & $5.29$ & $5.75$ & $6.87$ & $8.05$ & $10.20$ & $9.70$ & $10.69$ & $10.55$ & $11.08$ & $6.15$ & $8.29$\\
\hspace{1em}MinT(Shrink) & $5.00$ & $3.17$ & $2.55$ & $2.10$ & $2.72$ & $3.56$ & $4.65$ & $6.20$ & $5.55$ & $6.74$ & $6.85$ & $6.90$ & $3.17$ & $4.81$\\
\hspace{1em}EMinTU & $940.75$ & $1332.73$ & $1773.83$ & $2012.41$ & $2543.20$ & $2771.72$ & $3187.61$ & $3400.82$ & $3404.38$ & $3643.60$ & $3670.06$ & $3652.28$ & $1917.64$ & $2778.11$\\
\hspace{1em}IComb$[\alpha = 1, z_{x, y} = 1, c = 1]$ & $-7.48$ & $-7.12$ & $-5.72$ & $-6.41$ & $-5.75$ & $-7.83$ & $-7.23$ & $-7.56$ & $-6.90$ & $-5.38$ & $-4.69$ & $-3.65$ & $-6.72$ & $-6.23$\\
\hspace{1em}IComb$[\alpha = 1, z_{x, y} = 1, c = 0]$ & $\pmb{-8.07}$ & $\pmb{-8.11}$ & $-5.46$ & $\pmb{-7.18}$ & $-7.29$ & $\pmb{-9.75}$ & $\pmb{-9.21}$ & $\pmb{-9.28}$ & $-8.43$ & $-7.61$ & $-7.04$ & $-6.61$ & $\pmb{-7.66}$ & $-7.83$\\
\hspace{1em}IComb$[\alpha = 1, z_{x} = 1, c = 1]$ & $-4.08$ & $-1.94$ & $-4.04$ & $-4.87$ & $-6.40$ & $-8.23$ & $-8.72$ & $-7.78$ & $-9.25$ & $-6.17$ & $-7.06$ & $-7.66$ & $-4.99$ & $-6.48$\\
\hspace{1em}IComb$[\alpha = 1, z_{x} = 1, c = 0]$ & $-4.28$ & $-1.90$ & $-3.57$ & $-4.16$ & $-5.79$ & $-8.44$ & $-8.68$ & $-8.14$ & $\pmb{-9.68}$ & $-6.69$ & $-7.55$ & $-7.97$ & $-4.75$ & $-6.57$\\
\hspace{1em}IComb$[\alpha = 1, z_{x, y} = 0, c = 1]$ & $-4.44$ & $-2.63$ & $-3.12$ & $-4.06$ & $-5.17$ & $-6.18$ & $-6.23$ & $-5.96$ & $-7.13$ & $-5.95$ & $-6.15$ & $-6.19$ & $-4.30$ & $-5.37$\\
\hspace{1em}IComb$[\alpha = 1, z_{x, y} = 0, c = 0]$ & $-6.34$ & $-4.00$ & $-5.28$ & $-5.47$ & $-7.16$ & $-8.13$ & $-8.04$ & $-8.06$ & $-8.72$ & $-8.01$ & $-7.92$ & $-7.86$ & $-6.10$ & $-7.18$\\
\hspace{1em}IComb$[\alpha = 0, z_{x, y} = 1, c = 1]$ & $-4.98$ & $-5.65$ & $-5.08$ & $-5.48$ & $-5.06$ & $-6.08$ & $-6.08$ & $-4.68$ & $-4.88$ & $-3.12$ & $-2.75$ & $-2.83$ & $-5.39$ & $-4.63$\\
\hspace{1em}IComb$[\alpha = 0, z_{x, y} = 1, c = 0]$ & $-4.81$ & $-5.72$ & $-5.29$ & $-5.38$ & $-5.42$ & $-6.32$ & $-6.49$ & $-4.71$ & $-4.67$ & $-3.32$ & $-2.93$ & $-3.02$ & $-5.50$ & $-4.75$\\
\hspace{1em}IComb$[\alpha = 0, z_{x} = 1, c = 1]$ & $-5.11$ & $-5.57$ & $-5.04$ & $-5.54$ & $-5.11$ & $-6.09$ & $-6.08$ & $-4.51$ & $-4.85$ & $-3.12$ & $-2.77$ & $-2.70$ & $-5.42$ & $-4.61$\\
\hspace{1em}IComb$[\alpha = 0, z_{x} = 1, c = 0]$ & $-4.81$ & $-5.81$ & $-4.94$ & $-5.18$ & $-5.13$ & $-6.04$ & $-6.14$ & $-4.30$ & $-4.32$ & $-3.18$ & $-2.95$ & $-2.96$ & $-5.32$ & $-4.56$\\
\hspace{1em}IComb$[\alpha = 0, z_{x, y} = 0, c = 1]$ & $-6.20$ & $-4.68$ & $-5.08$ & $-5.42$ & $-6.70$ & $-7.27$ & $-7.76$ & $-7.79$ & $-8.85$ & $-7.50$ & $-7.91$ & $-7.85$ & $-5.91$ & $-7.01$\\
\hspace{1em}IComb$[\alpha = 0, z_{x, y} = 0, c = 0]$ & $-7.25$ & $-5.52$ & $\pmb{-6.24}$ & $-6.40$ & $\pmb{-7.79}$ & $-8.23$ & $-8.52$ & $-8.94$ & $-9.62$ & $\pmb{-8.51}$ & $\pmb{-8.98}$ & $\pmb{-8.73}$ & $-6.93$ & $\pmb{-7.99}$\\
\end{longtable}
\end{footnotesize}
\end{landscape}

\end{document}